\documentclass[manuscript]{aastex}
\usepackage{amsmath}
\usepackage{lscape}

\newcommand{\methanol}{CH$_3$OH}
\newcommand{\MF}{HCOOCH$_3$}
\newcommand{\hhcs}{H$_2$CS}
\newcommand{\Kzero}{$7_{0, 7}$--$6_{0, 6}$}
\newcommand{\Ktwo}{$7_{2, 5}$--$6_{2, 4}$}
\newcommand{\Kfour}{$7_{4, 4}$--$6_{4, 3}$/$7_{4, 3}$--$6_{4, 2}$}

\shorttitle{IRE in I16293A}
\shortauthors{Oya et al.}

\title{Infalling-Rotating Motion and Associated Chemical Change in the Envelope of IRAS 16293--2422 Source A Studied with ALMA}
\author{Yoko Oya\altaffilmark{1}, Nami Sakai\altaffilmark{2}, Ana L\'{o}pez--Sepulcre\altaffilmark{1}, Yoshimasa Watanabe\altaffilmark{1}, \\Cecilia Ceccarelli\altaffilmark{3, 4}, Bertrand Lefloch\altaffilmark{3, 4}, C\'{e}cile Favre\altaffilmark{3, 4}, Satoshi Yamamoto\altaffilmark{1}} 

\email{oya@taurus.phys.s.u-tokyo.ac.jp}
\altaffiltext{1}{Department of Physics, The University of Tokyo, 7-3-1, Hongo, Bunkyo-ku, Tokyo 113-0033, Japan}
\altaffiltext{2}{The Institute of Physical and Chemical Research (RIKEN), Wako, Saitama 351-0198, Japan}
\altaffiltext{3}{Universite Grenoble Alpes, IPAG, F-38000 Grenoble, France}
\altaffiltext{4}{CNRS, IPAG, F-38000 Grenoble, France}

\begin{abstract}
We have analyzed rotational spectral line emission of OCS, CH$_3$OH, HCOOCH$_3$, and H$_2$CS observed toward 
the low-mass Class 0 protostellar source IRAS 16293--2422 Source A at a sub-arcsecond resolution ($\sim$0\farcs6 $\times$ 0\farcs5) with ALMA. 
Significant chemical differentiation is found at a 50 AU scale. 
The OCS line is found to 
well trace the infalling-rotating envelope in this source. 
On the other hand, the CH$_3$OH and HCOOCH$_3$ distributions are found to be concentrated around the inner part of the infalling-rotating envelope. 
With a simple ballistic model of the infalling-rotating envelope, 
the radius of the centrifugal barrier (a half of the centrifugal radius) and the protostellar mass are evaluated from the OCS data to be from 40 to 60 AU and from 0.5 to 1.0 $M_\odot$, respectively, 
assuming the inclination angle of the envelope/disk structure to be 60\degr\ (90\degr\ for the edge-on configuration). 
Although the protostellar mass is correlated with the inclination angle, 
the radius of the centrifugal barrier is not. 
This is the first indication of the centrifugal barrier of the infalling-rotating envelope in a hot corino source.  
CH$_3$OH and HCOOCH$_3$ may be liberated from ice mantles due to weak accretion shocks 
around the centrifugal barrier, 
and/or due to 
protostellar heating. 
The H$_2$CS emission seems to come from the disk component inside the centrifugal barrier in addition to the envelope component. 
The centrifugal barrier 
plays a central role 
not only in the formation of a rotationally-supported disk but also in the 
chemical evolution from the envelope to the protoplanetary disk. 
\end{abstract}


\begin{document}
\section{Introduction} \label{sec:intro}
During the formation of solar-type stars, protostellar disks are formed around newly-born protostars, and they evolve into protoplanetary disks and planets. 
Hence, disk formation is an important subject in star-formation studies. 
Although the identification of rotationally-supported disks (Keplerian disks) has been reported in Class I sources and some Class 0 sources 
(Tobin et al. 2012; Yen et al. 2013; Murillo et al. 2013; Ohashi et al. 2014), 
it is still controversial when and how a disk structure is formed around a newly-born protostar. 
In addition, it is not clear how molecules in protostellar cores are processed during the disk formation and what kinds of molecules are finally delivered into the protoplanetary disks. 
These are key questions on chemical evolution from interstellar clouds to protoplanetary disks. 
Their thorough understanding will constrain the initial physical and chemical conditions of protoplanetary disks, 
which will eventually provide us important clues to the understanding of the origin of the Solar System.
It is thus essential to explore disk forming regions of various low-mass protostars with various molecular lines.

An important breakthrough toward understanding the disk formation and associated chemical evolution was recently reported by Sakai et al. (2014a) 
toward IRAS 04368+2557 in L1527 ($d$ = 137 pc) in the Cycle 0 operation of ALMA. 
This is a prototypical warm-carbon-chain-chemistry (WCCC) source, which harbors various carbon-chain molecules in the vicinity of the protostar (Sakai et al. 2008; Sakai \& Yamamoto 2013). 
Sakai et al. (2014a, b) discovered the centrifugal barrier of the infalling-rotating envelope at a radius of 100 AU from the protostar 
based on observations of CCH, c-C$_3$H$_2$, and CS lines. 
A simple ballistic model of the infalling-rotating envelope (Oya et al. 2014, 2015) can well explain the kinematic structure observed in these lines. 
At the centrifugal barrier, not only the gas motion is discontinuous, 
but also the chemical composition of the gas drastically changes (Sakai et al. 2014b). 
Carbon-chain related molecules (CCH, c-C$_3$H$_2$) and CS disappear inward of the centrifugal barrier, 
while SO (and possibly CH$_3$OH) is enhanced around the centrifugal barrier and exists inside it. 
CCH, c-C$_3$H$_2$ and CS are thought to be depleted onto dust grains due to long stay in the dense mid-plane region ($n$(H$_2$) $\sim 10^8$ cm$^{-3}$) just inward of the centrifugal barrier. 
On the other hand, SO (and possibly CH$_3$OH) seems to be liberated from dust grains due to weak accretion shocks in front of the centrifugal barrier. 
Thus, chemical compositions highlight the centrifugal barrier. 
It is most likely that a disk structure is being formed inside the centrifugal barrier in this source. 
The existence of a rotationally-supported disk is indeed reported by the observation of CO isotopologue lines (Tobin et al. 2012; Ohashi et al. 2014). 

The centrifugal barrier appears as a natural consequence of the angular momentum conservation and the energy conservation of the infalling gas, 
and its radius is determined by the specific angular momentum and the protostellar mass.  
It is related to a transition zone from the infalling-rotating envelope to the rotationally-supported disk, 
and hence, it should exist in other protostars.  
Nevertheless, its identification has been difficult by observations of the spectral lines of CO and its isotopologues, 
which have long been employed to trace the kinematic structure of protostellar sources: 
The emission of these lines comes from the entire region including the infalling-rotating envelope and the rotationally-supported disk.  
Observations of various molecular species at a high spatial resolution with ALMA have opened a new avenue 
to identify and characterize the infalling-rotating envelope and its centrifugal barrier by using the drastic changes in chemical composition. 
In order to apply such chemical diagnostics to many other sources, we need to know what molecular species traces what part of the protostellar source.  
In the L1527 case, the carbon-chain molecules and SO are a good combination to highlight the centrifugal barrier, as mentioned above (Sakai et al. 2014a, b).  
Here, we investigate the chemical and kinematic structures of the envelope gas toward IRAS 16293--2422, 
whose chemical composition is known to be quite different from L1527 (e.g. Caux et al. 2011; J\o rgensen et al. 2011; Sakai \& Yamamoto 2013).  

IRAS 16293--2422 is a well-known Class 0 protostar in Ophiuchus ($d$ = 120 pc; Knude \& H\o g, 1998), 
whose molecular gas distribution and dynamics, outflows, and chemical composition have been extensively studied 
(e.g. Wootten 1989; Mizuno et al. 1990; Mundy et al. 1990, 1992; Blake et al. 1994; van Dishoeck et al. 1995; Looney et al. 2000; 
Ceccarelli et al. 2000a, b, c; Sch\"{o}ier et al. 2002; Cazaux et al. 2003; Bottinelli et al. 2004; Takakuwa et al. 2007; 
Yeh et al. 2008; Caux et al. 2011; J\o rgensen et al. 2011). 
While L1527 is rich in carbon-chain molecules, 
IRAS 16293--2422 is rich in complex organic molecules (COMs) such as HCOOCH$_3$ and (CH$_3$)$_2$O in the vicinity of the protostars 
(e.g. Sch\"{o}ier et al. 2002; Cazaux et al. 2003; Bottinelli et al. 2004; Kuan et al. 2004; Pineda et al. 2012). 
This characteristic chemical composition is known as hot corino chemistry (Ceccarelli 2004; Bottinelli et al. 2004). 
IRAS 16293--2422 consists of Source A and Source B, which are separated by $5\arcsec$ (Chandler et al. 2005). 
The protostellar masses of Source A and Source B are reported to be $\sim 1\ M_\odot$ and $< 0.1\ M_\odot$, respectively, 
and it is suggested that Source B rotates around Source A (Bottinelli et al. 2004; Caux et al. 2011). 
Since Source B itself has an envelope/disk component with a face-on configuration, 
its rotating motion can scarcely be detected (Pineda et al. 2012; Zapata et al. 2013). 
Inverse P Cygni profiles are instead reported for the COM lines toward Source B, 
indicating that COMs are in the gas infalling to the protostar (Pineda et al. 2012; Zapata et al. 2013). 
On the other hand, Source A shows a rotation signature (Pineda et al. 2012). 
Favre et al. (2014) recently reported the observation of C$^{34}$S ($J$=7--6) with SMA and eSMA at a resolution of $0\farcs46 \times 0\farcs29$, 
and revealed a clear spin-up feature toward the protostar. 
They discussed two possibilities to explain the origin of the rotating motion: a Keplerian disk and an infalling-rotating envelope. 
Since CS line is known to trace the infalling-rotaing envelope in L1527 (Oya et al. 2015), 
it is likely that the rotation signature found by Favre et al. (2014) is due to the infalling-rotating envelope around IRAS 16293--2422 Source A. 
In this study, we have analyzed ALMA archival data of other molecular species 
to examine whether the centrifugal barrier of the infalling-rotating envelope can be identified in this source. 

\section{Data} \label{sec:obs}
The archival data of IRAS 16293--2422 observed in the ALMA Cycle 1 operation are used in this paper. 
We analyzed the spectral lines listed in Table \ref{tb:lines}. 
Images were obtained by using the CLEAN algorithm. 
The details of the observation are described in Appendix \ref{sec:data_ALMA}. 
The continuum image was prepared by averaging line-free channels, 
and the line maps were obtained after subtracting the continuum component directly from the visibilities. 
The synthesized-beam sizes for the continuum and the spectral lines are listed in Table \ref{tb:lines}. 
The rms noise levels for the continuum, OCS, CH$_3$OH, HCOOCH$_3$, and H$_2$CS maps are 2.0, 2.0, 4.0, 1.8, and 2.0 mJy beam$^{-1}$, respectively, 
for a channel width of 122 kHz. 
The coordinates of Source A are derived from a 2D Gaussian-fit at the continuum peak: 
($\alpha_{2000}$, $\delta_{2000}$) = ($16^{\rm h} 32^{\rm m} 22\fs8713 \pm 0\fs0012$, $-24\degr 28\arcmin 36\farcs5023 \pm 0\farcs0111$). 
In this paper, we focus only on Source A to examine its kinematic structure, 
because the rotating signature is seen in Source A but not in Source B as mentioned in Section \ref{sec:intro}. 

\section{Line intensity Distribution} \label{sec:distribution}
\begin{landscape}
\begin{figure}
	\includegraphics[bb = 0 100 100 550, scale = 0.55]{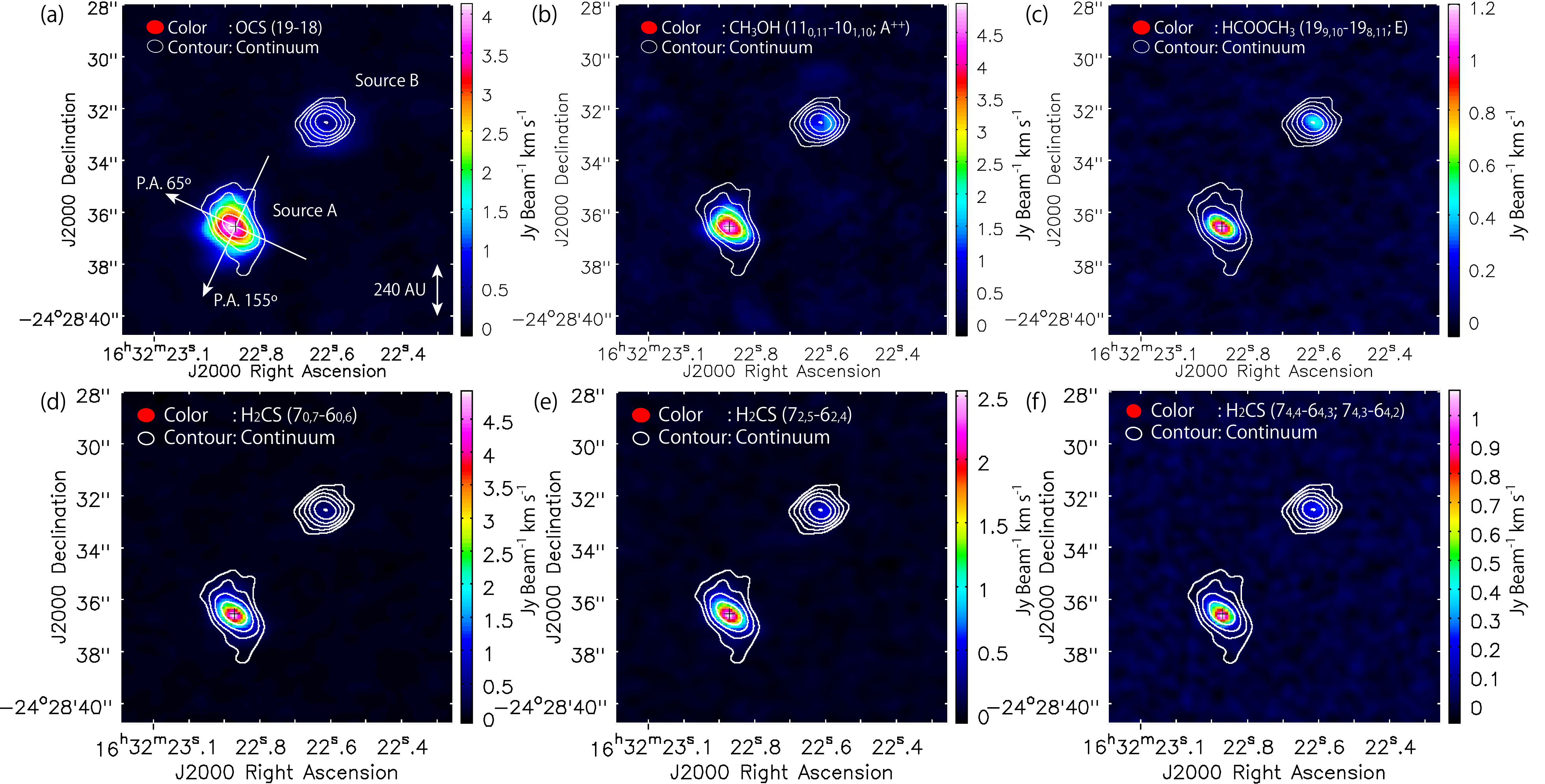}
	\vspace*{40pt}
	\caption{Moment 0 maps (integrated intensity maps) of OCS (a), CH$_3$OH (b), HCOOCH$_3$ (c), and H$_2$CS (d--f) (color). 
			The integrated velocity ranges are from $-6.1$ to 14.0, from $-5.9$ to 14.1, from $-11.2$ to 18.8, from $-7.3$ to 20.9, from $-7.3$ to 20.9, and from $-13.2$ to 11.2 km s$^{-1}$ 
			for panels (a), (b), (c), (d), (e), and (f), respectively. 
			White contours represent the continuum emission at 1.2 mm. 
			The contour levels for the continuum are 10, 20, 40, 80, 160, and 320$\sigma$, 
			where $\sigma =$ 2 mJy beam$^{-1}$. 
			White arrows in panel (a) represent the envelope direction (P.A. 65\degr) and the outflow direction (P.A. 155\degr) along which the PV diagrams 
			in Figures \ref{fig:PV2_obs} and \ref{fig:PV2_H2CS-Contour}--\ref{fig:PV2_H2CS_model} are prepared. 
			The position of Source A is represented by black crosses: ($\alpha_{2000}$, $\delta_{2000}$) = ($16^{\rm h} 32^{\rm m} 22\fs8713$, $-24\degr 28\arcmin 36\farcs5023$). 
			\label{fig:moment0}}
\end{figure}
\end{landscape}

The ALMA archival data of IRAS 16293--2422 (Cycle 1) contain a number of spectral lines of various molecules including COMs. 
In order to trace the infalling-rotating envelope suggested by the C$^{34}$S line observed with SMA and eSMA (Favre et al. 2014), 
we focused on the molecular line whose distribution is the most extended. 
We found that the OCS line ($J$=19--18) is the best for this purpose. 
Figure \ref{fig:moment0}(a) shows the moment 0 map of OCS. 
The 2D Gaussian-fitted size of the OCS line intensity distribution deconvolved by the beam is $(1\farcs665 \pm 0\farcs017) \times (1\farcs351 \pm 0\farcs015)$ (P.A. $21\degr\!\!.2 \pm 2\degr\!\!.2$) in FWHM. 
This size is almost comparable to that of C$^{34}$S reported by Favre et al. (2014), 
whose deconvolved size is estimated to be 1\farcs6 by using the SMA/eSMA archival data. 
Hence, it is most likely that OCS also traces the infalling-rotating envelope. 
This prediction will be verified by the analysis of its velocity structure in the following sections. 
It should be noted that the peak of the C$^{34}$S distribution is offset by $0\farcs52$ from the dust continuum peak 
(submillimeter continuum source Aa reported by Chandler et al. 2005), 
while such an offset is smaller in OCS ($0\farcs14$) than in C$^{34}$S. 
In contrast, the CH$_3$OH (11$_{0, 11}$--10$_{1, 10}$; A$^{++}$) and HCOOCH$_3$ (19$_{9, 10}$--19$_{8, 11}$; E) distributions, 
whose moment 0 maps are shown in Figures \ref{fig:moment0}(b) and \ref{fig:moment0}(c), respectively, 
are more compact and more centrally concentrated than OCS and C$^{34}$S. 
Figures \ref{fig:moment0}(d--f) show the moment 0 maps of the three lines of H$_2$CS (\Kzero; \Ktwo; \Kfour). 
They are also concentrated around the protostar. 
The 2D Gaussian-fitted sizes of the CH$_3$OH, HCOOCH$_3$, and H$_2$CS (\Kzero) intensitiy distributions deconvolved by the beam are 
$(1\farcs042 \pm 0\farcs018) \times (0\farcs862 \pm 0\farcs016)$ (P.A. $46\degr\!\!.4 \pm 4\degr\!\!.5$), 
$(0\farcs728 \pm 0\farcs023) \times (0\farcs371 \pm 0\farcs020)$ (P.A. $58\degr\!\!.5 \pm 2\degr\!\!.5$), 
and $(0\farcs876 \pm 0\farcs011) \times (0\farcs572 \pm 0\farcs007)$ (P.A. $79\degr\!\!.3 \pm 1\degr\!\!.1$) in FWHM, respectively. 

\section{Velocity Structure} \label{sec:vel}
\subsection{OCS} \label{sec:vel_OCS}
Figure \ref{fig:moment1}(a) shows the moment 1 map of OCS. 
It reveals a rotation signature around the protostar Source A; 
a clear velocity gradient is found in the northeast-southwest direction, 
which is consistent with that previously found by Takakuwa et al. (2007), Rao et al. (2009), Pineda et al. (2012), and Favre et al. (2014).  
Here, we employ 3.8 km s$^{-1}$ as the systemic velocity of Source A, 
judging from the range of the systemic velocity in the previous reports (3.6 km s$^{-1}$; Takakuwa et al. 2007: 3.9 km s$^{-1}$: Bottinelli et al. 2004: 3.8 km s$^{-1}$; Favre et al. 2014). 
In order to define the position angle of the envelope/disk system, 
we investigated the peak positions of the distributions of the most red-shifted emission ($v_{\rm lsr}$ = 10.4--10.7 km s$^{-1}$) 
and the most blue-shifted emission ($v_{\rm lsr}$ = $-2.8$-- $-3.1$ km s$^{-1}$) of OCS. 
These two components have peaks with slight offsets from the continuum peak position, 
and the continuum and the emission peaks are well on a common line. 
The position angle of the disk/envelope system is thus determined to be 65\degr$\pm$10\degr, as shown in Figure \ref{fig:moment0}(a). 
Then, the position angle of the rotation axis (i.e., outflow direction) is 155\degr$\pm$10\degr, 
being consistent with the previous report (P.A. 144\degr; Favre et al. 2014). 

Figures \ref{fig:PV2_obs}(a) and \ref{fig:PV2_obs}(b) show the position-velocity (PV) diagrams of OCS, 
where the position axes in panels (a) and (b) are along 
the disk/envelope direction (referred to hereafter as envelope direction) 
and the direction perpendicular to it (referred to hereafter as outflow direction), respectively, 
as shown in Figure \ref{fig:moment0}(a). 
The PV diagram of OCS along the envelope direction (Figure \ref{fig:PV2_obs}a) shows a clear spin-up feature toward the protostar. 
The PV diagram of OCS along the outflow direction (Figure \ref{fig:PV2_obs}b) shows a significant velocity gradient. 
Since the OCS emission is concentrated around the protostar, the contribution of outflows is unlikely. 
This velocity gradient most likely represents an infalling motion of the rotating envelope in the vicinity of the protostar. 
Such a velocity gradient along the outflow direction is similar to that of the infalling motion observed for CS in L1527 (Oya et al. 2015). 
Hence, the rotating motion traced by OCS is not Keplerian, 
but the gas is infalling and rotating at a scale of 400 AU in diameter. 
This situation is schematically illustrated in Figure \ref{fig:schema} (right). 

\subsection{CH$_3$OH and HCOOCH$_3$} \label{sec:vel_COMs}
\begin{landscape}
\begin{figure}
	\includegraphics[bb = 0 100 500 500, scale =0.32]{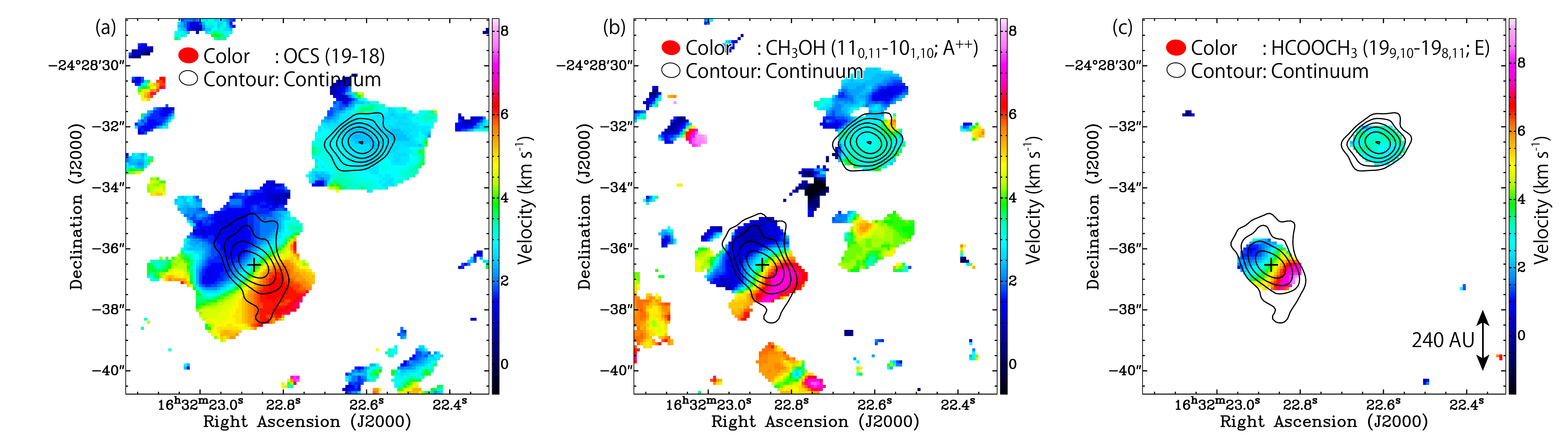}
	\vspace*{40pt}
	\caption{Moment 1 maps of OCS (a), CH$_3$OH (b), and HCOOCH$_3$ (c) (color). 
			Black contours represent the continuum at 1.2 mm, 
			which are the same as the white contours in Figure \ref{fig:moment0}. 
			The position of Source A is represented by black crosses. 
			\label{fig:moment1}}
\end{figure}
\end{landscape}

\begin{figure}
	\includegraphics[bb = -60 30 300 1500, scale = 0.37]{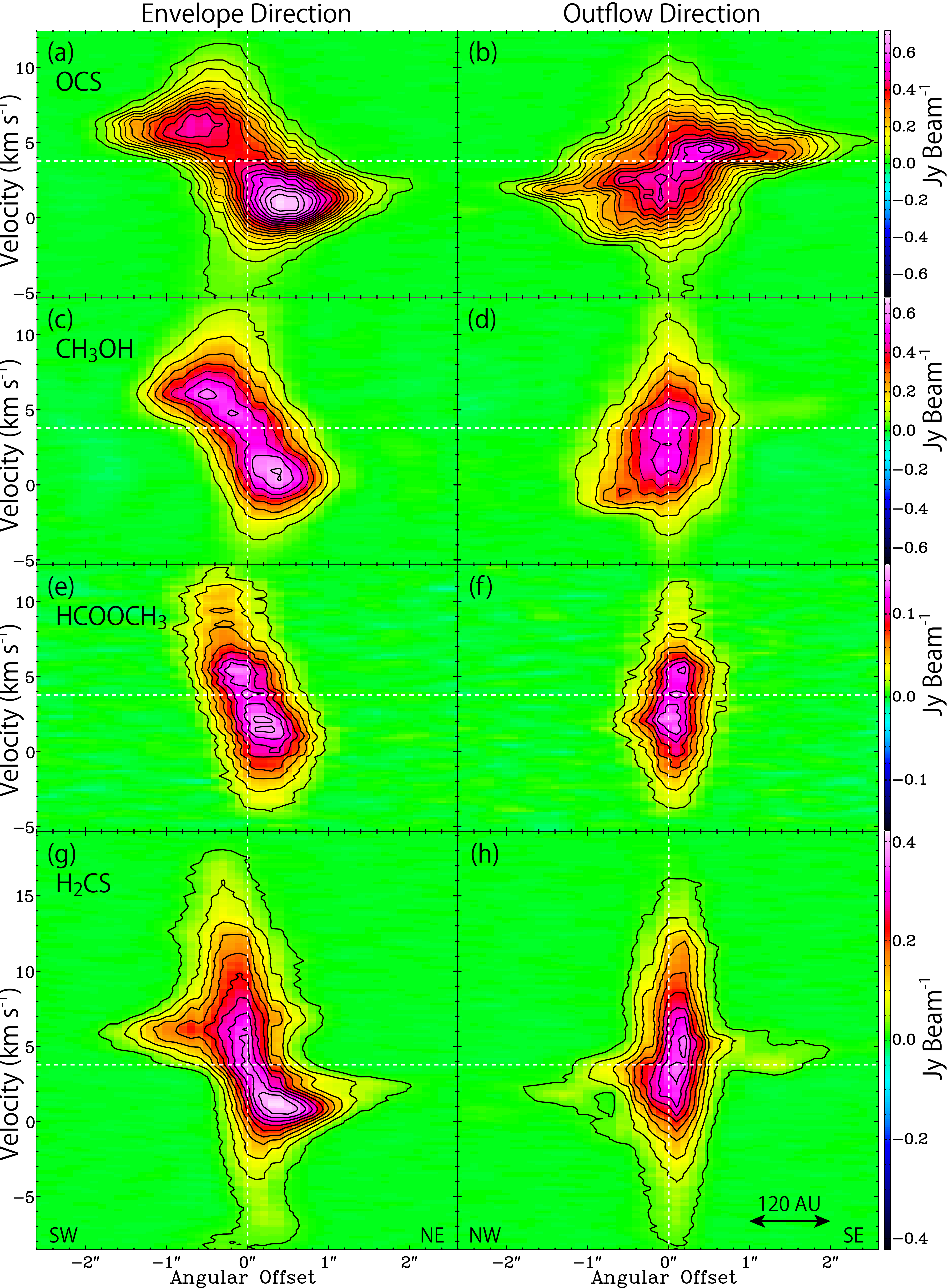}
	\caption{PV diagrams of OCS (a, b), CH$_3$OH (c, d), HCOOCH$_3$ (e, f) and H$_2$CS (\Kzero; g, h) 
	along the envelope direction (P.A. 65\degr) and the outflow direction (P.A. 155\degr) shown in Figure \ref{fig:moment0}(a). 
			 The first contour levels are at 20$\sigma$, 20$\sigma$, 10$\sigma$, and 10$\sigma$ 
			 and the level steps are 20$\sigma$, 20$\sigma$, 10$\sigma$, and 20$\sigma$ for OCS, \methanol, \MF, and \hhcs, 
			 where $\sigma$ = 2.0, 4.0, 1.8, and 2.0 mJy beam$^{-1}$, respectively. 
			\label{fig:PV2_obs}}
\end{figure}

\begin{figure}
	\includegraphics[bb = 0 0 100 300, scale = 0.74]{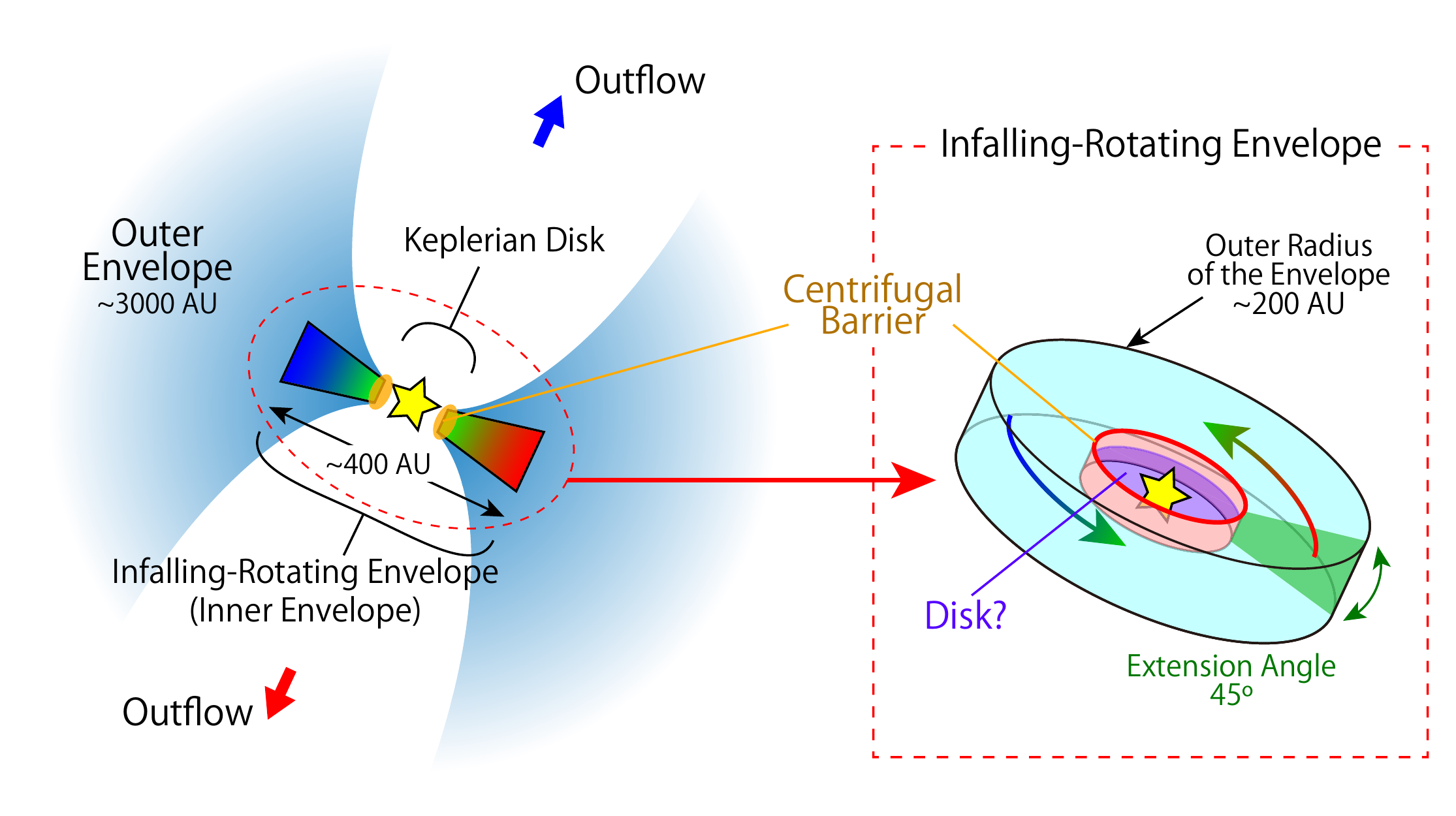}
	\caption{Schematic illustration of the gas components around Source A. 
			The outer envelope part represented by light blue regions in the left figure is resolved out in this observation. 
			The kinematic structure in the inner envelope component, which is assumed to be cylindrical around the protostar, 
			is analyzed with our infalling-rotating envelope model in Section \ref{sec:IREmodel}. 
			The Keplerian disk component, which is the most inner part in the figure and is inside the centrifugal barrier of the infalling-rotating envelope, 
			is traced by \hhcs\ (Figures \ref{fig:PV2_obs}g, h). 
			Its kinematic structure is discussed in Section \ref{sec:model_H2CS}. 
			\label{fig:schema}}
\end{figure}

Figure \ref{fig:moment1}(b) depicts the moment 1 map of CH$_3$OH, 
which clearly shows a velocity gradient around the protostar. 
The PV diagram of CH$_3$OH along the envelope direction (Figure \ref{fig:PV2_obs}c) also supports the rotation signature, as in the case of OCS. 
However, the distribution is more concentrated near the protostar than that of OCS (Figure \ref{fig:PV2_obs}a), as mentioned in Section \ref{sec:distribution}. 
The PV diagram of CH$_3$OH along the outflow direction (Figure \ref{fig:PV2_obs}d) shows little velocity gradient unlike the OCS case (Figure \ref{fig:PV2_obs}b), 
suggesting less infalling motion. 
This also suggests that the distribution is rather concentrated around a more inner part of the infalling-rotating envelope than that of OCS, 
because the infalling velocity vanishes when approaching the centrifugal barrier of the infalling-rotating envelope (Sakai et al. 2014a; Oya et al. 2014). 
In the PV diagram of CH$_3$OH along the envelope direction (Figure \ref{fig:PV2_obs}c), 
the intensity peaks seem to appear at the $+ 0\farcs5$ and $- 0\farcs5$ positions with the blue-shifted and red-shifted velocity, respectively. 
This feature can be explained if the emitting region of CH$_3$OH mainly exists in 
a ring-like structure with an apparent radius of about 0\farcs5 in the innermost part of the infalling-rotating envelope. 

Figure \ref{fig:moment1}(c) shows the moment 1 map of HCOOCH$_3$, 
while Figures \ref{fig:PV2_obs}(e) and \ref{fig:PV2_obs}(f) are the PV diagrams of HCOOCH$_3$ along the envelope and outflow directions, respectively. 
The velocity structure of the HCOOCH$_3$ line is essentially similar to that of CH$_3$OH, 
and is more concentrated around the protostar. 
Although the PV diagram of CH$_3$OH along the envelope direction (Figure \ref{fig:PV2_obs}c) reveals a slight extension toward the southwestern side, 
such a feature is absent in HCOOCH$_3$ (Figure \ref{fig:PV2_obs}e). 
No velocity gradient along the outflow direction is seen (Figure \ref{fig:PV2_obs}f), 
suggesting that the HCOOCH$_3$ emission mainly comes from the gas without infalling motion. 
This feature is similar to the case seen in SO toward L1527 and TMC-1A (Sakai et al. 2014a, b, 2016). 

\subsection{H$_2$CS} \label{sec:vel_H2CS}
\begin{figure}
	\includegraphics[bb = 0 50 1000 1600, scale = 0.37]{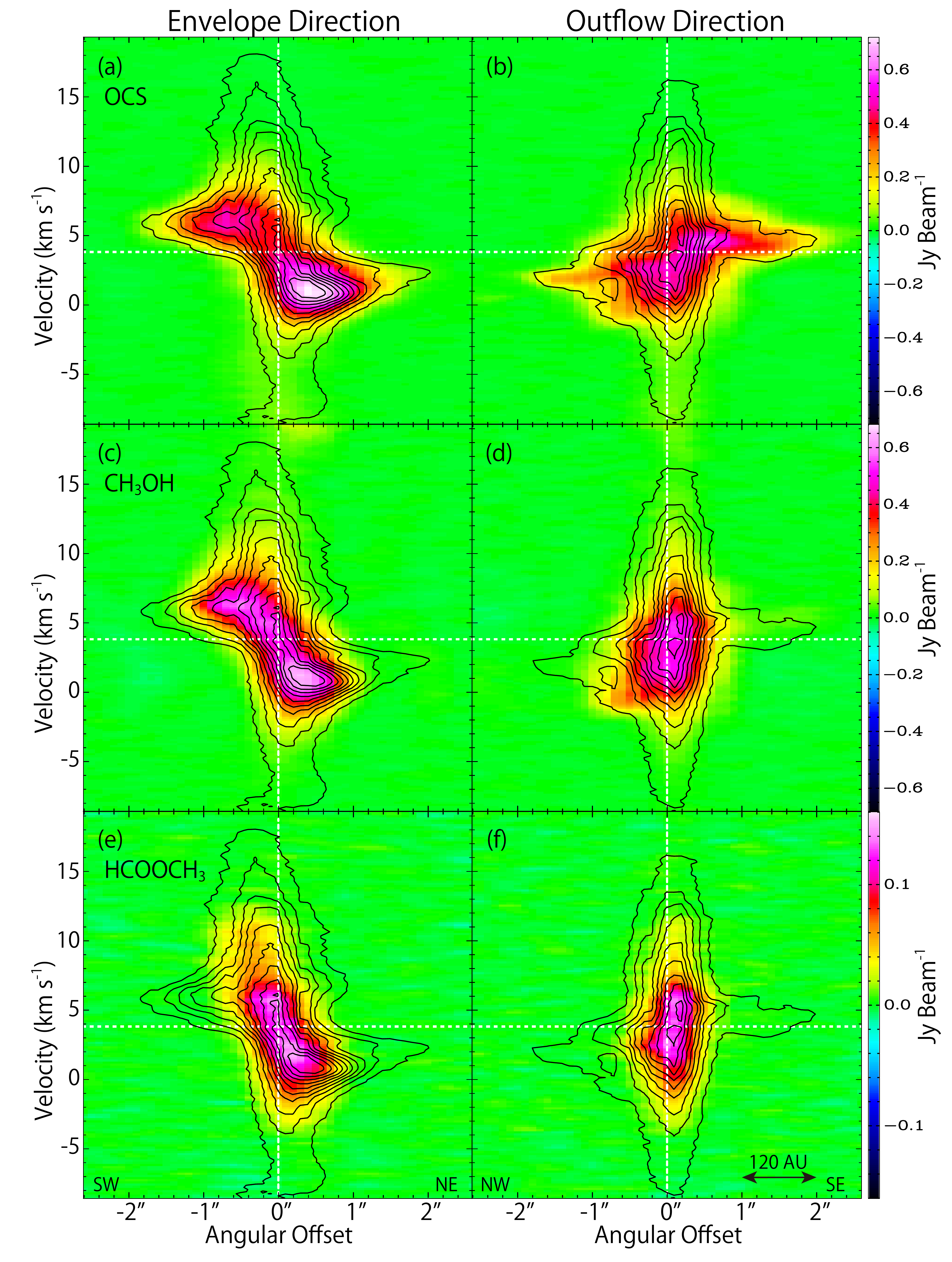}
	\caption{PV diagrams of OCS, \methanol, \MF\ (color) and H$_2$CS (\Kzero; black contours) along the envelope direction (P.A. 65\degr) and the outflow direction (P.A. 155\degr). 
			The contour levels for H$_2$CS are every 20$\sigma$ from 10$\sigma$, where $\sigma$ = 2.0 mJy beam$^{-1}$. 
			\label{fig:PV2_H2CS-Contour}}
\end{figure}

Figures \ref{fig:PV2_obs}(g) and \ref{fig:PV2_obs}(h) show the PV diagrams of H$_2$CS (\Kzero) along the envelope and outflow directions. 
Figure \ref{fig:PV2_H2CS-Contour} also shows them in contours superposed on those of OCS, \methanol, and \MF\ in color. 
In the moment 0 maps (Figures \ref{fig:moment0}d--f), the distribution of the H$_2$CS emission 
looks concentrated around the protostar as in the case of CH$_3$OH and HCOOCH$_3$ (Figures \ref{fig:moment0}b, c). 
However, the PV diagrams are much different from those of these two molecules (Figure \ref{fig:PV2_H2CS-Contour}). 
First, the PV diagrams of \hhcs\ contain broader velocity components near the protostar. 
For instance, the maximum velocity-shift from the systemic velocity (3.8 km s$^{-1}$) is as large as 14 km s$^{-1}$, 
which is about twice the corresponding velocity shift ($\sim$7 km s$^{-1}$) seen in the OCS line (Figures \ref{fig:PV2_H2CS-Contour}a, b). 
The broad-velocity components seen in H$_2$CS may trace the Keplerian disk component in the vicinity of the protostar. 
In addition, weak emission from the envelope component is also visible in the PV diagram of \hhcs\ along the envelope direction 
unlike the \methanol\ and \MF\ cases (Figures \ref{fig:PV2_H2CS-Contour}c, e), 
which is very similar to the distribution of OCS, as shown in Figure \ref{fig:PV2_H2CS-Contour}a. 
Hence, H$_2$CS seems to reside in all the regions from the infalling-rotating envelope to the disk component inside it. 
Such a behavior of H$_2$CS is similar to that seen in H$_2$CO toward L1527 (Sakai et al. 2014b); 
H$_2$CO resides in all the regions from the envelope to the disk component in L1527. 

\section{Infalling-Rotating Envelope Model} \label{sec:IREmodel}
As described above, the different molecules allow us to trace different parts of the gas around the protostar. 
In this section, we analyze their distributions and kinematic structures by using a simple physical model. 
Although the envelope gas of IRAS 16293--2422 is reported to extend on a 3000 AU scale and to have a spherical configuration (Ceccarelli et al. 2000a), 
we here focus on the small scale ($\sim 400$ AU) structure showing the rotation motion of the inner envelope and the disk around the protostar, 
as illustrated in Figure \ref{fig:schema}. 
The spherical outer envelope component is resolved out in this observation, 
as shown in the moment 0 maps (Figure \ref{fig:moment0}). 
Hence we assume a cylindrical model to examine the compact and flattened structure in the vicinity of the protostar.

\subsection{OCS} \label{sec:model_OCS}
The velocity structure of the OCS line (Figures \ref{fig:PV2_obs}a, \ref{fig:PV2_obs}b) suggests the existence of an infalling-rotating envelope around the dust continuum peak. 
Hence, we analyze it with a simple model of an infalling-rotating envelope (Oya et al. 2014), 
which has been applied to L1527, IRAS 15398--3359, and TMC-1A (Sakai et al. 2014a, b, 2016; Oya et al. 2014, 2015). 
In our model, the gas motion is approximated by the particle motion. 
Since we are interested in the velocity structure, we assume a density profile proportional to $r^{-1.5}$ and a constant abundance of molecules for simplicity. 
We also assume optically thin emission, and do not consider radiative transfer and excitation effects. 
Although this model is a simplified one, it well reproduces the velocity structures of L1527 and TMC-1A (Sakai et al. 2014a, b, 2016; Oya et al. 2015). 
The velocity field in the infalling-rotating envelope can be determined by the two parameters: 
the radius of the centrifugal barrier ($r_{\rm CB}$) and the protostellar mass ($M$) (Sakai et al. 2014a; Oya et al. 2014). 
The model includes additional parameters to account for the observed PV diagrams: 
the inclination angle of the disk/envelope system, 
the extending angle of the envelope thickness, and the outer radius of the envelope ($R$), as shown in Figure \ref{fig:schema}. 
Note that the outer radius does not mean the size of the envelope, but the size of the distribution of molecular emission. 
The model image is convolved by a line width of 1 km s$^{-1}$ 
and the synthesized beam of ($0\farcs65 \times 0\farcs51$; P.A. 85.\!\!\degr29) for the OCS line (Table \ref{tb:lines}). 
This line width is assumed by considering the possible turbulent motions in the inner envelope. 
 
\begin{figure}
	\includegraphics[bb = 0 50 100 1700, scale = 0.28]{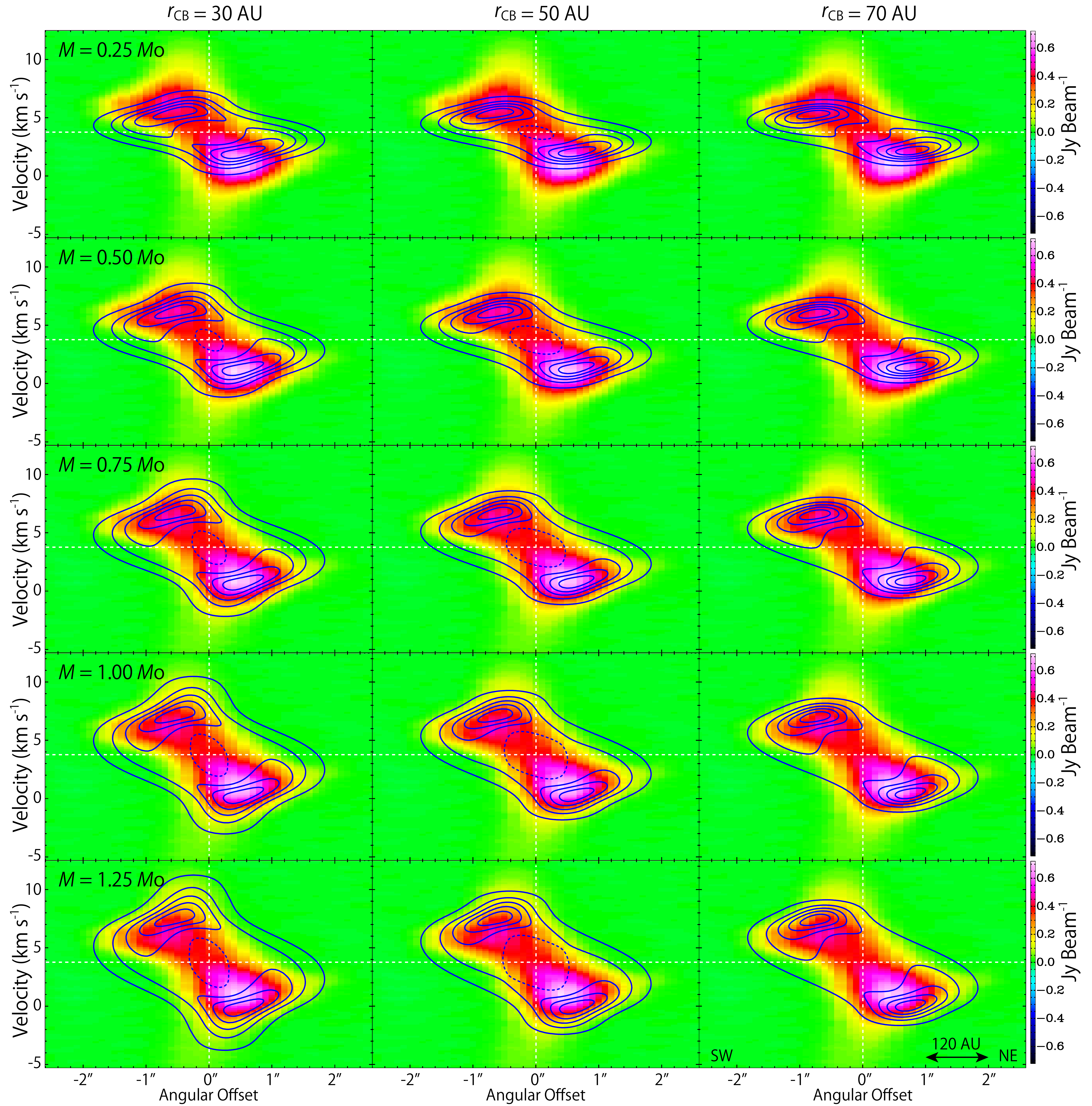}
	\caption{Results of the model simulations (blue contours) with various sets of the radius of the centrifugal barrier ($r_{\rm CB}$) and the protostellar mass ($M$) 
			superposed on the PV diagram of OCS (color) along the envelope direction (P.A. 65\degr). 
			The other physical parameters for the models are set as follows; 
			$i = 60\degr$ and $R = 180$ AU. 
			The line width is assumed to be 1 km s$^{-1}$. 
			The contour levels are every 20 \% from 5 \% of each peak intensity. 
			The dashed contours around the central position in the panels for $r_{\rm CB}$ = 30 and 50 AU 
			represent the dip toward the center. 
			\label{fig:PV_varParams_envelope}}
\end{figure}

\begin{figure}
	\includegraphics[bb = 0 50 100 1700, scale = 0.28]{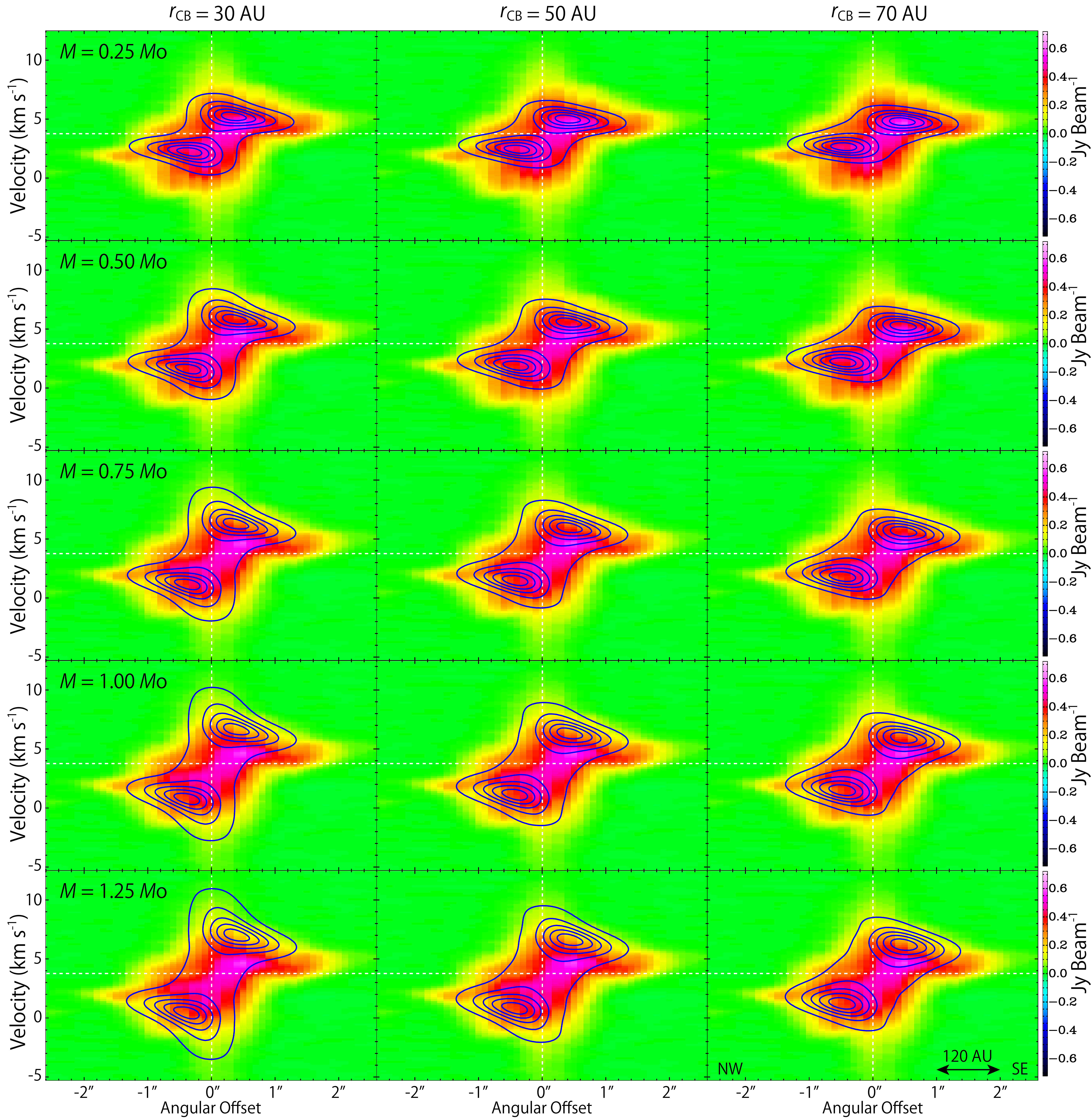}
	\caption{Results of the model simulations (blue contours) with various sets of the radius of the centrifugal barrier ($r_{\rm CB}$) and the protostellar mass ($M$) 
			superposed on the PV diagram of OCS (color) along the outflow direction (P.A. 155\degr). 
			The other physical parameters and the contour levels are the same as those in Figure \ref{fig:PV_varParams_envelope}. 
			\label{fig:PV_varParams_outflow}}
\end{figure}

We conducted simulations of the PV diagrams of OCS with a wide range of parameters. 
As an example, simulations of the PV diagram along the envelope direction 
with various sets of the protostellar mass and the radius of the centrifugal barrier are shown in Figure \ref{fig:PV_varParams_envelope}, 
while those along the outflow direction in Figure \ref{fig:PV_varParams_outflow}. 
These figures show how sensitive to the protostellar mass and the radius of the centrifugal barrier the simulated PV diagrams are. 
Judging from the goodness of the simultaneous fit for the two directions (Figures \ref{fig:PV_varParams_envelope} and \ref{fig:PV_varParams_outflow}) by eye, 
a protostellar mass of 0.75 $M_\odot$ and a radius of the centrifugal barrier of 50 AU reasonably reproduce the PV diagrams, 
assuming an inclination angle of 60\degr\ (90\degr\ for the edge-on configuration). 
To verify this result,  
we calculated the root mean square (rms) of the residuals (Table \ref{tb:goodness}), 
and the rms value is confirmed to be almost the lowest for the above parameters. 
We note that the rms values of the residuals in Table \ref{tb:goodness} themselves do not have a statistical meaning, 
because we employ a simple model concentrating on the velocity structure of the envelope 
and do not consider abundance variation, asymmetrical distribution of molecules, optical depth effects, and so on. 
Systematic errors caused by these assumptions would overwhelm the statistical noise, 
which makes the chi-square analysis difficult (see also Oya et al. 2015). 
Nevertheless, it should be stressed that even such a simple model can reasonably explain the basic velocity structure observed in the OCS line. 

\begin{figure}
	\includegraphics[bb = -80 50 300 1500, scale = 0.36]{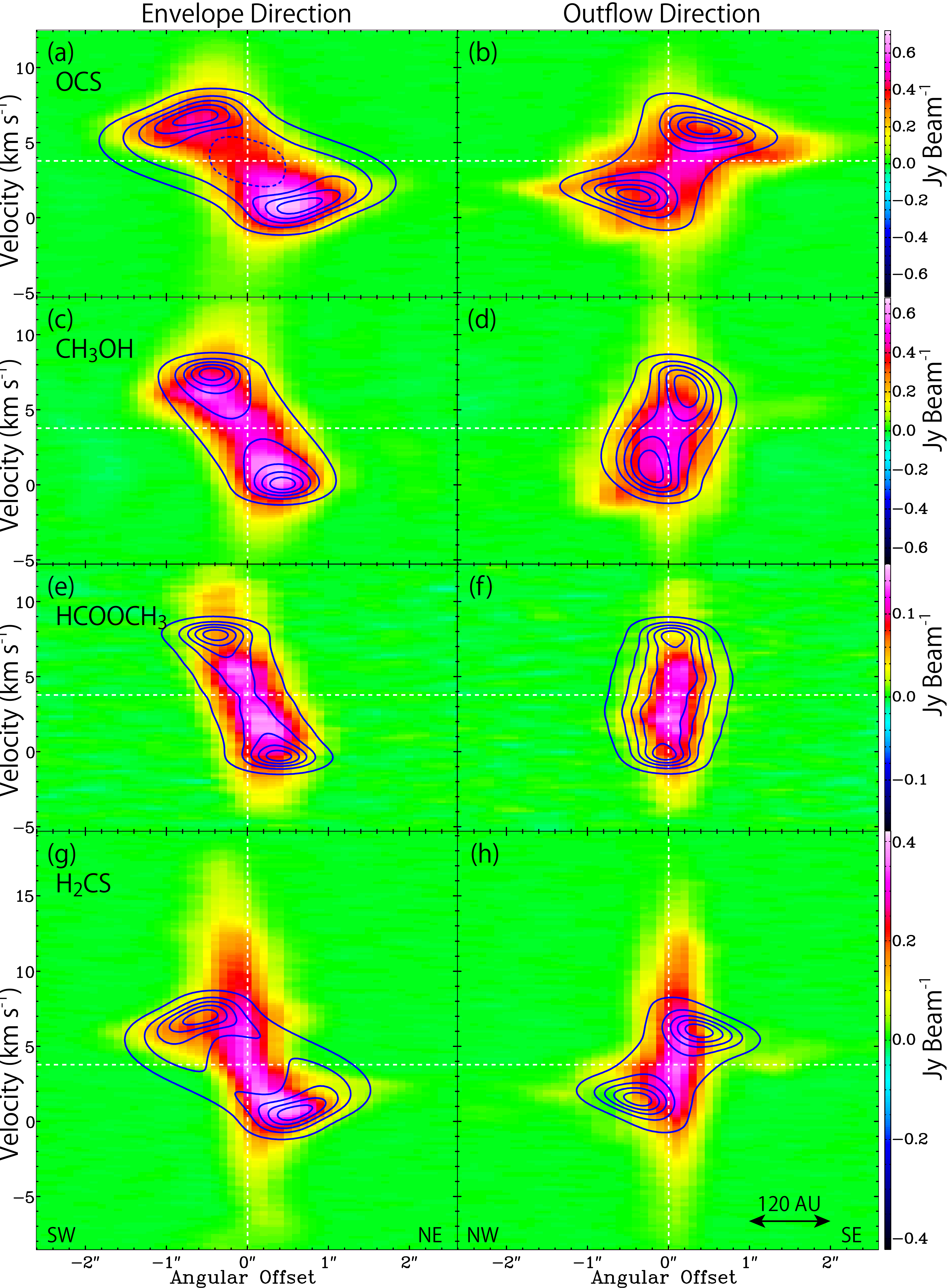}
	\caption{Color maps are the same as those in Figure \ref{fig:PV2_obs}. 
			Blue contours represent the results of the infalling-rotating envelope models, 
			where $M = 0.75\ M_\odot$, $r_{\rm CB} = 50$ AU, and $i = 60\degr$. 
			The outer radii of the envelope $R$ in the models for OCS, CH$_3$OH, HCOOCH$_3$, and \hhcs\ are 180, 80, 55, and 150 AU, respectively. 
			The contour levels are every 20 \% from 5 \% of each peak intensity. 
			Meaning of the dashed contour in panel (a) is denoted in footnotes of Figure \ref{fig:PV_varParams_envelope}. 
			\label{fig:PV2_model}}
\end{figure}

Further comparisons were made by considering the other model parameters including the inclination angle, 
which allowed us to finally constrain the model parameters. 
As a result, the radius of the centrifugal barrier lies in the range from 40 AU to 60 AU. 
The outer radius of the OCS distribution is derived to be 180 AU. 
The results of the simulation for the PV diagrams of OCS 
along the envelope and outflow directions for the inclination angle of 60\degr\ 
are shown in Figures \ref{fig:PV2_model}(a) and \ref{fig:PV2_model}(b). 

It is important to note that the protostellar mass is highly correlated with the inclination angle of the envelope, 
while the radius of the centrifugal barrier is not. 
Indeed, an acceptable agreement between the observation and the model result is obtained 
for an inclination angle ranging between 30\degr\ and 70\degr. 
For a given radius of the centrifugal barrier ($r_{\rm CB}$), 
the protostellar mass ($M$) which reproduces the maximum velocity-shift due to the rotation (4.5 km s$^{-1}$) 
depends on the inclination angle ($i$) as: 
\begin{equation}
	\frac{M}{M_\odot} = 0.57 \left( \frac{r_{\rm CB}}{50\ {\rm AU}}\right) \sin^{-2} i, 
\end{equation}
where $i =$ 90\degr\ for an edge-on configuration. 
For instance, the protostellar mass is 0.97 $M_\odot$ and 0.64 $M_\odot$ for an inclination angle of 50\degr\ and 70\degr, respectively, for $r_{\rm CB}$ of 50 AU. 
The protostellar mass is estimated by Bottinelli et al. (2004) and Caux et al. (2011) 
to be $\sim 1\ M_\odot$ from the difference in the $v_{\rm lsr}$ between Source A and Source B 
and the velocity width of the observed lines, respectively, 
with which our result is almost consistent. 

\begin{figure}
	\includegraphics[bb = -60 70 100 1950, scale = 0.26]{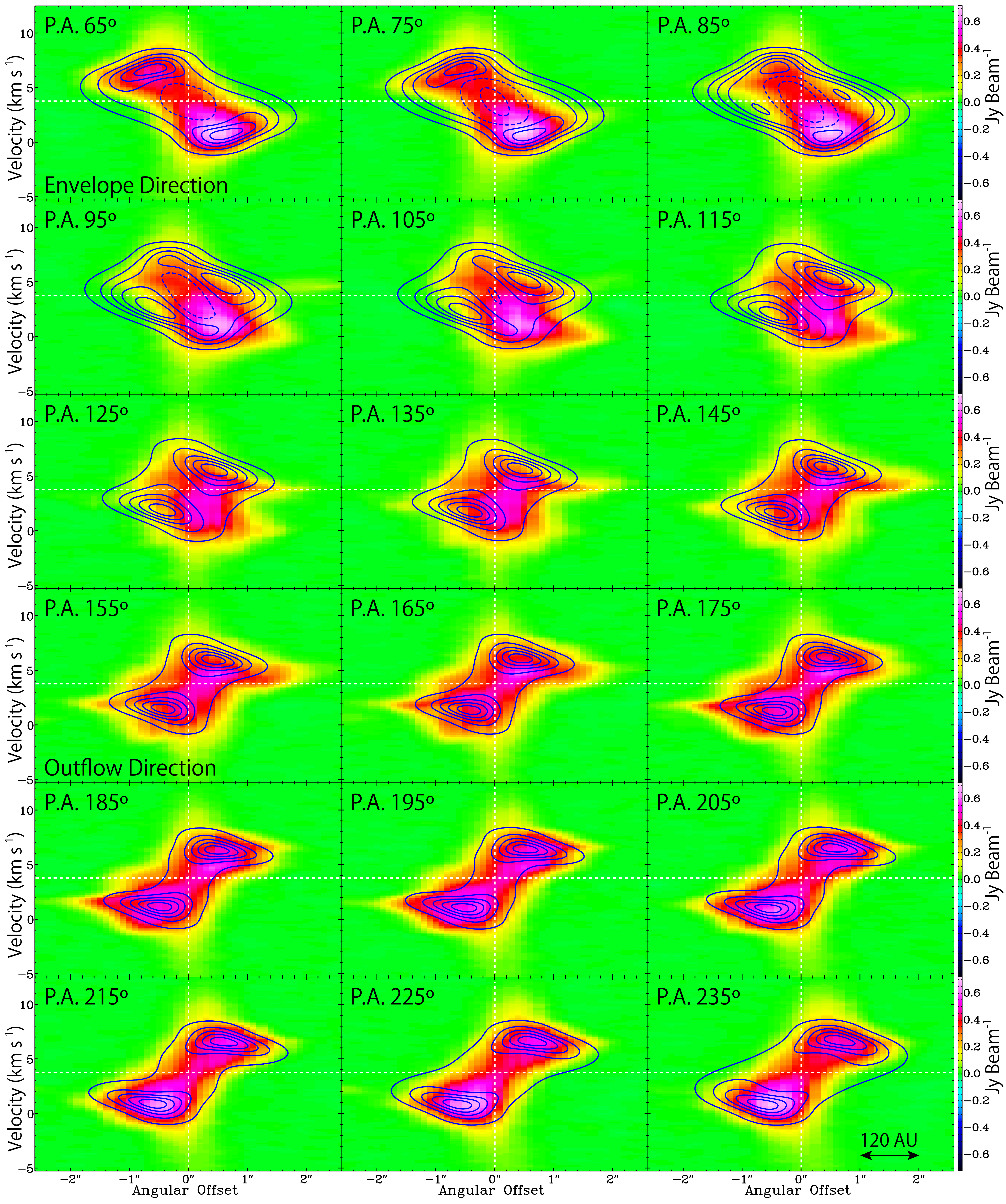}
	\caption{PV diagrams of OCS (color) for the 18 lines passing through the protostar position with various position angles. 
			The position angles of the lines along which the PV diagrams are prepared are every 10\degr\ from the envelope direction (P.A. 65\degr). 
			Blue contours represent the results of the infalling-rotating envelope model, 
			where $M = 0.75\ M_\odot$, $r_{\rm CB} = 50$ AU, $i = 60\degr$, and $R = 180$ AU. 
			The line width is assumed to be 1 km s$^{-1}$. 
			The contour levels are every 20 \% from 5 \% of each peak intensity. 
			Meaning of the dashed contours in panels `P.A. 65\degr', `P.A. 75\degr', `P.A. 85\degr', `P.A. 95\degr', and `P.A. 105\degr' is denoted in footnotes of Figure \ref{fig:PV_varParams_envelope}. 
			\label{fig:PV18_OCS}}
\end{figure}

The PV diagrams of OCS along the lines passing through the protostar position for every 10\degr\ of the position angle are shown in Figure \ref{fig:PV18_OCS}. 
Starting from the outflow direction (P.A. 155\degr), 
features of the PV diagrams do not change symmetrically between clockwise rotation and counter-clockwise rotation of the position angles. 
This behavior cannot be explained by the Keplerian motion, but requires both rotation and infalling motions. 
The model basically reproduces the trend of the PV diagrams for the various position angles, 
although we can still find some discrepancies in detailed distributions. 
The discrepancies seem to originate from asymmetry of the molecular distribution in the infalling-rotating envelope around the protostar, 
as seen in the moment 0 map (Figure \ref{fig:moment0}a). 
For instance, the extension of the gas is narrowest along the 125\degr\ (P.A.) line instead of 155\degr\ (P.A.; outflow direction). 
This asymmetry makes the discrepancy between the observation and the simulation larger around that position angle (Figure \ref{fig:PV18_OCS}). 
Another reason for this discrepancy would be the relative contribution of the rotation and infalling motions. 
Since these two motions have almost opposite directions along the line of sight 
for the position angle from 75\degr\ to 145\degr, 
the PV diagram is sensitive to a small difference of their relative contribution. 
This situation makes difficult to reproduce the observed PV diagrams for these position angles. 

\begin{figure}
	\includegraphics[bb = 0 0 100 100, scale = 0.41]{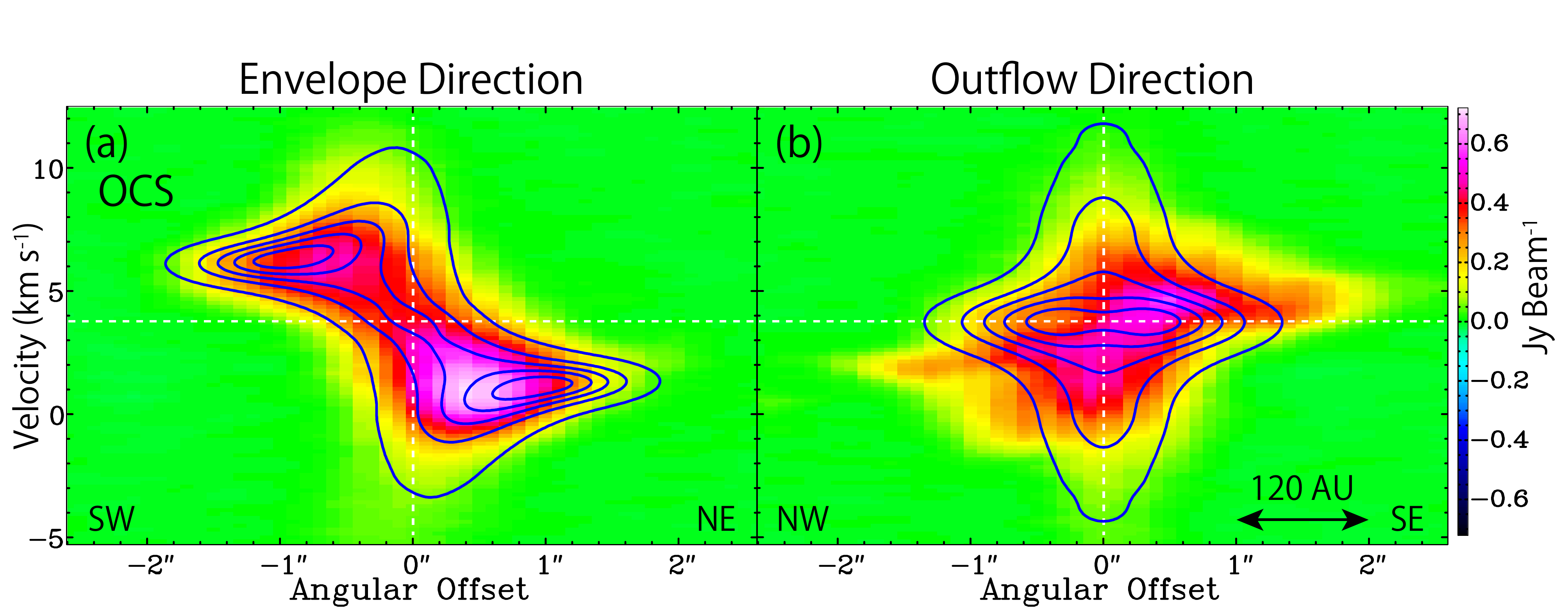}
	\caption{PV diagrams of OCS (color) along the envelope direction (P.A. 65\degr) and the outflow direction (P.A. 155\degr). 
			Blue contours represent the results of the Keplerian motion with a protostellar mass of 1.5 $M_\odot$, 
			where the outer radius of the model is set to be 180 AU. 
			The contour levels are every 20 \% from 5 \% of each peak intensity. 
			\label{fig:PV2_OCS_Kep}}
\end{figure}

It should be stressed that the observed PV diagrams for OCS cannot be explained only by the Keplerian motion, as mentioned above. 
Figure \ref{fig:PV2_OCS_Kep} shows the simulation of the PV diagrams along the envelope and outflow directions assuming the Keplerian motion. 
While the PV diagram along the envelope direction can be explained to some extent, 
the velocity gradient along the outflow direction cannot be reproduced. 
Nevertheless, we cannot rule out the possibility that OCS may also reside in the Keplerian disk to some extent in addition to the infalling-rotating envelope, 
which contributes to the shape of the PV diagrams. 
Neglecting this contribution may cause additional systematic errors in the above analysis. 

In this model, we assume an $r^{-1.5}$ density profile and a constant abundance of molecules for simplicity. 
In order to assess how this assumption affects the results, 
we also conducted the simulations by using density profiles of $r^0$ and $r^{-2.5}$. 
These may partly account for the effect of the optical depth, the excitation effect, and the temperature gradient effect in an effective way. 
However, we found that the simulation results of the PV diagrams do not change significantly. 

Finally, we note that the PV diagrams of the infalling-rotating envelope model 
can also explain the essential part of the observed PV diagrams of C$^{34}$S along the envelope 
observed with SMA and eSMA (Favre et al. 2014), 
by using $M$ and $r_{\rm CB}$ derived for OCS. 

\subsection{CH$_3$OH and HCOOCH$_3$} \label{sec:model_COMs}
\begin{figure}
	\includegraphics[bb = -30 50 100 1950, scale = 0.27]{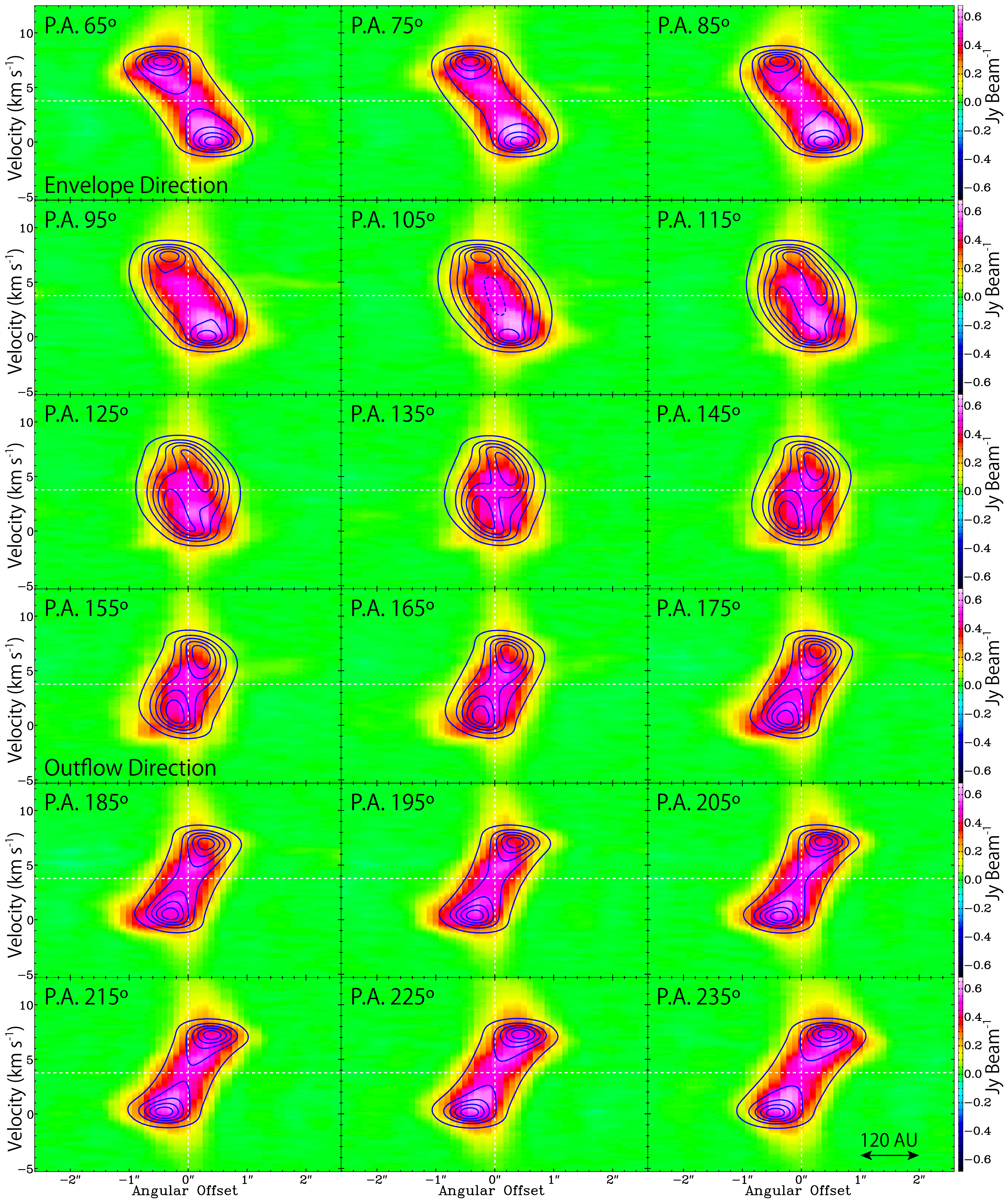}
	\caption{PV diagrams of CH$_3$OH (color) for the 18 lines passing through the protostar position with various position angles. 
			The position axes are the same as those in Figure \ref{fig:PV18_OCS}. 
			Blue contours represent the results of the infalling-rotating envelope model, 
			where $M = 0.75\ M_\odot$, $r_{\rm CB} = 50$ AU, $i = 60\degr$, and $R = 80$ AU. 
			The line width is assumed to be 1 km s$^{-1}$. 
			The contour levels are every 20 \% from 5 \% of each peak intensity. 
			Meaning of the dashed contour in panel `P.A. 105\degr' is denoted in footnotes of Figure \ref{fig:PV_varParams_envelope}. 
			\label{fig:PV18_CH3OH}}
\end{figure}

\begin{figure}
	\includegraphics[bb = -30 50 100 1950, scale = 0.27]{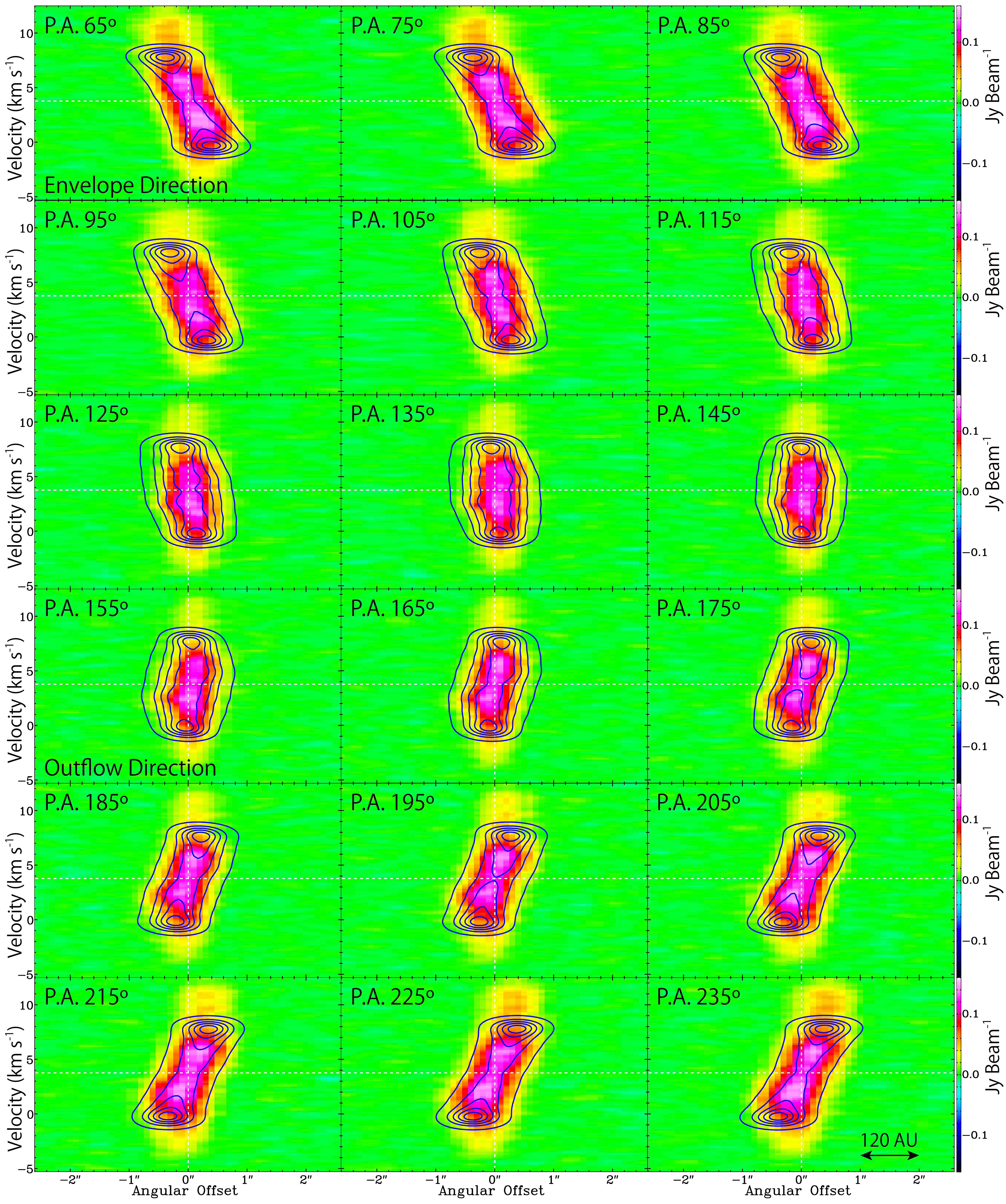}
	\caption{PV diagrams of HCOOCH$_3$ (color) for the 18 lines passing through the protostar position with various position angles. 
			The position axes are the same as those in Figure \ref{fig:PV18_OCS}. 
			Blue contours represent the results of the infalling-rotating envelope model, 
			where $M = 0.75\ M_\odot$, $r_{\rm CB} = 50$ AU, $i = 60\degr$, and $R = 55$ AU. 
			The line width is assumed to be 1 km s$^{-1}$. 
			The contour levels are every 20 \% from 5 \% of each peak intensity. 
			\label{fig:PV18_HCOOCH3}}
\end{figure}

We also compare the PV diagrams of CH$_3$OH and HCOOCH$_3$ along the envelope and outflow directions 
with the model results in Figures \ref{fig:PV2_model}(c--f). 
The model parameters are set to be the same as those derived from the analysis of OCS 
($M = 0.75\ M_\odot$, $r_{\rm CB} = 50$ AU, and $i = 60\degr$), except for the outer radius, 
and the model images are convolved by a line width of 1 km s$^{-1}$ and the synthesized beam for each line listed in Table \ref{tb:lines}. 
As mentioned in Section \ref{sec:distribution}, 
the distributions of CH$_3$OH and HCOOCH$_3$ are compact and their 2D Gaussian-fitted sizes deconvolved by the beam are 
130 AU $\times$ 100 AU and 90 AU $\times$ 40 AU in FWHM, respectively, 
assuming a distance of 120 pc (Knude \& H\o g, 1998). 
Hence, the outer radii of the infalling-rotating envelope model for \methanol\ and \MF\ are set to be 80 AU and 55 AU, respectively. 
The latter value just assumes that \MF\ is mostly distributed in a ring-like structure at the centrifugal barrier. 
The PV diagrams simulated with the small outer radii show a good agreement with the observations (Figures \ref{fig:PV2_model}c--f). 
Figures \ref{fig:PV18_CH3OH} and \ref{fig:PV18_HCOOCH3} show a more detailed comparison using the PV diagrams for every 10\degr\ of the position angle. 
Although there are some discrepancies between the observation and the model results, 
such as the missing of high velocity-shift components in \MF, 
the model reasonably reproduces the observed PV diagrams. 
Hence, the emitting regions of CH$_3$OH and HCOOCH$_3$ seem to essentially have a ring-like structure around the centrifugal barrier. 
Such distributions of CH$_3$OH and HCOOCH$_3$ are similar to the SO ring around the protostar in L1527 (Sakai et al. 2014a, b). 
Note that the outer radius employed above for CH$_3$OH ($R$ = 80 AU) is close to the centrifugal radius (twice the radius of the centrifugal barrier) by chance, 
where the centrifugal force balances with the gravitational force. 

\subsection{H$_2$CS} \label{sec:model_H2CS}
\begin{figure}
	\includegraphics[bb = -260 60 100 1530, scale = 0.32]{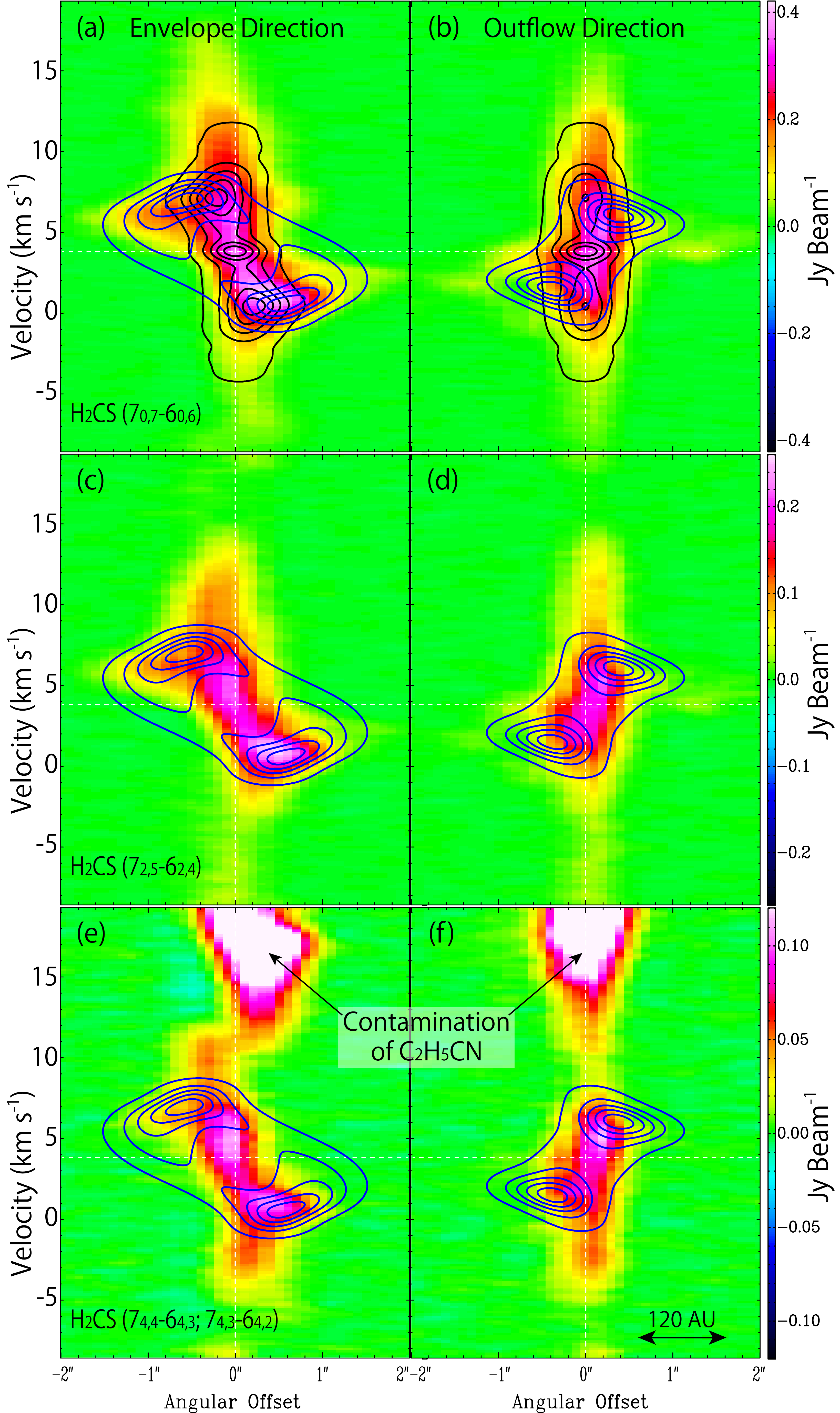}
	\caption{PV diagrams of H$_2$CS (\Kzero; \Ktwo; \Kfour) along the envelope direction (P.A. 65\degr) and the outflow direction (P.A. 155\degr). 
			Blue contours represent the results of the infalling-rotating envelope model, 
			where $M = 0.75\ M_\odot$, $r_{\rm CB} = 50$ AU, $i = 60\degr$, and $R = 150$ AU. 
			The line width is assumed to be 1 km s$^{-1}$. 
			Black contours in panels (a) and (b) represent the results of the simulation for the Keplerian motion inside the centrifugal barrier, 
			where $M = 0.75\ M_\odot$ and $i = 60\degr$. 
			The contour levels for the two models are every 20 \% from 5 \% of each peak intensity. 
			The H$_2$CS ($7_{4, 4}$--$6_{4, 3}$ and $7_{4, 3}$--$6_{4, 2}$) line is contaminated 
			by the C$_2$H$_5$CN ($28_{1, 28}$--$27_{1, 27}$; 240.3193373 GHz) line in panels (e) and (f). 
			\label{fig:PV2_H2CS_model}}
\end{figure}

In Figure \ref{fig:PV2_H2CS_model}, we overlay the results of the model simulation on the observed PV diagrams of the three lines of H$_2$CS (\Kzero; \Ktwo; \Kfour). 
In this simulation, we used the same parameters as those used for OCS ($M = 0.75\ M_\odot$, $r_{\rm CB} = 50$ AU, and $i = 60\degr$) 
except for a smaller outer radius of the envelope ($R$ = 150 AU). 
The model images are convolved by a line width of 1 km s$^{-1}$ and the synthesized beam for \hhcs\ (\Kzero) shown in Table \ref{tb:lines}. 
The simulation result reproduces the observed PV diagrams for the envelope part, 
although the distribution of the \Kfour\ line tends to be less extended to the envelope than the other two lines as in the case of \methanol\ and \MF. 

On the other hand, the components observed near the protostar apparently show broader velocity widths than the results of the infalling-rotating envelope model. 
These broader velocity components seem to trace the Keplerian disk component in the vicinity of the protostar. 
Hence, we made a simulation of the PV diagram for the Keplerian disk assuming the protostellar mass 
determined from the above analysis of the infalling-rotating envelope in Section \ref{sec:model_OCS} (0.75 $M_\odot$). 
Here, we also assume that the outer radius of the Keplerian disk is equal to the radius of the centrifugal barrier (50 AU) for simplicity. 
The result of this disk model is also overlaid on the observed PV diagrams for \hhcs\ (\Kzero) in Figures \ref{fig:PV2_H2CS_model}(a) and \ref{fig:PV2_H2CS_model}(b). 
It well explains the broader-velocity components, which strongly suggests the existence of the Keplerian disk structure within the centrifugal barrier. 

\section{Discussion} \label{sec:discuss}
\subsection{Infalling-Rotating Envelope and Its Centrifugal Barrier} \label{sec:discuss_CB}
The results presented above most likely show the existence of the infalling-rotating envelope and the indication of its centrifugal barrier 
in the prototypical hot corino source IRAS 16293--2422 Source A. 
The observed PV diagrams are reasonably explained by the simple ballistic model of the infalling-rotating envelope 
(Figures \ref{fig:PV2_model}, \ref{fig:PV18_OCS}, \ref{fig:PV18_CH3OH}--\ref{fig:PV2_H2CS_model}). 
The existence of the centrifugal barrier is not shown as clearly as in the case of L1527 (Sakai et al. 2014a) 
because of the contamination by the Keplerian disk component in this source. 
Nevertheless, we can estimate its radius 
by using the ballistic model. 
For the identification of the centrifugal barrier in the hot corino source, 
the emission of OCS is found to be useful.  
L1527 and IRAS 16293--2422 are different from each other in the chemical composition, 
but their physical structures are found to be similar. 
Hence, the infalling-rotating envelope and its centrifugal barrier would exist in protostellar sources regardless of their chemical characteristics.  

On the other hand, the radius of the centrifugal barrier is different from source to source (Sakai et al. 2014a, b, 2016; Oya et al. 2014, 2015). 
The range of the centrifugal barrier is from a few tens of AU to 100 AU in radius. 
According to the ballistic model (Oya et al. 2014), 
the size of the centrifugal barrier is proportional to the square of the specific angular momentum brought into the infalling-rotating envelope, 
and is inversely proportional to the protostellar mass. 
This means that the radius of the centrifugal barrier is sensitive to the environment of the parent core 
where the spatial distribution of the specific angular momentum of the gas is determined. 
The magnitude of the specific angular momentum will also affects disk structures and outflows. 
Statistical studies of the radius of the centrifugal barrier for a number of young protostellar sources are therefore interesting, 
and will provide us with rich information regarding the formation process of protoplanetary disks and their diversity. 

\subsection{Origin of the Chemical Change around the Centrifugal Barrier} \label{sec:discuss_chem}
The present analyses demonstrate chemical differentiation in the closest vicinity of the protostar in the hot corino source. 
A chemical change seems to be occurring in the region around the centrifugal barrier, 
probably between the centrifugal radius and the centrifugal barrier, 
as in the case of the WCCC sources L1527 and TMC-1A (Sakai et al. 2014b, 2016). 
In IRAS 16293--2422 Source A, the OCS emission is mainly distributed in the infalling-rotating envelope up to the radius of 180 AU, 
while the \methanol\ and \MF\ emission is mainly concentrated around the centrifugal barrier. 
On the other hand, \hhcs\ resides from the envelope to the Keplerian disk. 

So far, OCS has been detected in hot cores and hot corinos (Blake et al., 1987, 1994; Caux et al. 2011). 
It is known to be abundant in IRAS 16293--2422, 
because high-excitation lines of the isotopologues, OC$^{34}$S and O$^{13}$CS, are detected by single-dish observations (Blake et al. 1994; Caux et al. 2011). 
According to the jump-model analysis by Sch\"{o}ier et al. (2002), its fractional abundance relative to H$_2$ is as high as 10$^{-7}$. 
However, production processes of OCS are not well understood. 
The gas phase production (CS + OH, SO + CH) as well as the solid phase production are proposed (Wakelam et al. 2011; Loison et al. 2012). 
The evaporation temperature of OCS is evaluated to be 60 K from the surface binding energy of 2888 K 
(UMIST database for astrochemistry; McElroy et al. 2013, http://udfa.ajmarkwick.net/index.php), 
and is lower than the gas temperature of the infalling-rotating envelope derived below from the multiple line analysis of \hhcs. 
Hence the liberation of OCS from dust grains cannot be ruled out for the distribution in the infalling-rotating envelope. 

On the other hand, the CH$_3$OH and HCOOCH$_3$ emission is distributed in a narrow region (possibly a ring-like structure) around the centrifugal barrier.  
Concentration of these molecules around the centrifugal barrier suggests 
that they would be liberated from ice mantles to the gas phase due to weak accretion shocks expected in front of the centrifugal barrier. 
This situation is similar to the SO distribution in the WCCC sources L1527 and TMC-1A (Sakai et al. 2014a, b, 2016). 
Alternatively, they would be thermally evaporated by protostellar heating within a certain radius from the protostar. 
If this evaporation radius is just outside the radius of the centrifugal barrier by chance, 
the observed distributions of \methanol\ and \MF\ can be explained. 
Since the luminosity of IRAS 16293--2422 is as high as 22 $L_\odot$ (Crimier et al. 2010), 
the CH$_3$OH evaporation region would be able to extend outward of the centrifugal barrier. 
According to the spherical model by Crimier et al. (2010), the gas kinetic temperature at the radius of 50 AU is estimated to be as high as 130 K,  
and the outer radius of the infalling-rotating envelope model of 80 AU for CH$_3$OH coincides with the water sublimation radius ($\gtrsim$ 100 K). 
Hence thermal evaporation by the protostellar heating can cause such an enhancement of CH$_3$OH and HCOOCH$_3$. 

In order to assess these two possibilities, we derived the gas kinetic temperature by using the \Kzero, \Ktwo, \Kfour\ lines of H$_2$CS. 
These three transitions can be used as a good thermometer, as shown below, 
because the cross $K$-ladder radiative transitions (i.e. $K_a$ = 2 $\rightarrow$ 0) are very slow. 
Since H$_2$CS is distributed from the infalling-rotating envelope to the Keplerian disk component, 
we derived the gas kinetic temperatures of the infalling-rotating envelope component, the centrifugal barrier, and the Keplerian disk (Table \ref{tb:temperature}). 
These temperatures are derived from the integrated intensity ratios of the \Ktwo\ and \Kfour\ lines to the \Kzero\ line. 
The integrated intensities are calculated for a circular area of the moment 0 maps with a diameter of 0\farcs5, 
which is centered at the distance of 1\farcs0, 0\farcs5, and 0\farcs0 from the continuum peak along the envelope direction (Figure \ref{fig:moment0}a) 
for the envelope component, the centrifugal barrier, and the Keplerian disk component, respectively. 
This diameter is comparable to the size of the synthesized beam. 
The velocity-shift range for the integration is from 1.0 to 4.0 km s$^{-1}$, from 2.0 to 5.0 km s$^{-1}$, and from 5.0 to 7.3 km s$^{-1}$, 
for the blue-shifted and red-shifted parts of the envelope component, the centrifugal barrier, and the Keplerian disk component, respectively. 
Here the systemic velocity component is excluded, because it can be affected by self-absorption. 
For the Keplerian disk component, only the velocity-shift range higher than the maximum rotation velocity of 
the infalling-rotating envelope is used so as to exclude the contamination by the envelope component. 

\begin{figure}
	\includegraphics[bb = 0 0 100 900, scale = 0.26]{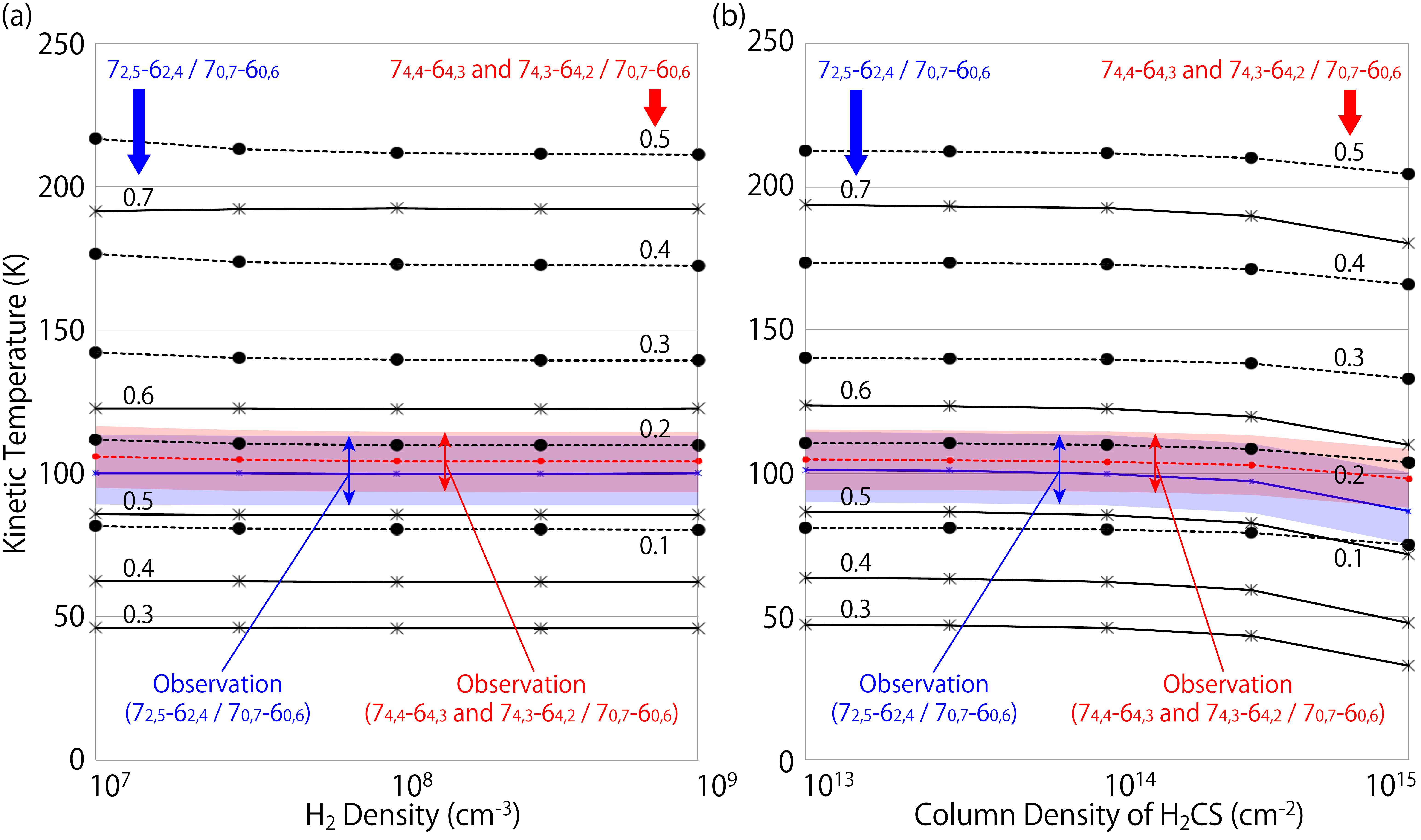}
	\caption{An example of determination of the gas kinetic temperature by using the intensity ratios of 
			the H$_2$CS ($7_{2, 5}$--$6_{2, 4}$; $7_{4, 4}$--$6_{4, 3}$ and $7_{4, 3}$--$6_{4, 2}$) lines relative to the H$_2$CS ($7_{0, 7}$--$6_{0, 6}$) line. 
			Solid and dashed lines represent the gas kinetic temperature for a given ratio of 
			$7_{2, 5}$--$6_{2, 4}$/$7_{0, 7}$--$6_{0, 6}$ and ($7_{4, 4}$--$6_{4, 3}$ and $7_{4, 3}$--$6_{4, 2}$)/$7_{0, 7}$--$6_{0, 6}$, respectively, 
			as a function of the H$_2$ density (a) and the \hhcs\ column density (b) 
			calculated by using RADEX code (van der Tak et al. 2007). 
			The column density of \hhcs\ is fixed to be $10^{14}$ cm$^{-2}$ in panel (a), 
			while the H$_2$ density is fixed to be $10^8$ cm$^{-3}$ in panel (b). 
			The solid blue and dashed red lines represent the observed results of 
			the $7_{2, 5}$--$6_{2, 4}$/$7_{0, 7}$--$6_{0, 6}$ and ($7_{4, 4}$--$6_{4, 3}$ and $7_{4, 3}$--$6_{4, 2}$)/$7_{0, 7}$--$6_{0, 6}$ ratios 
			in the red-shifted envelope component, respectively, 
			and the blue colored and red colored areas represent their error ranges (3$\sigma$). 
			The derived ranges of the gas kinetic temperature for the other components are listed in Table \ref{tb:temperature}. 
			The error is estimated only from the statistical error 
			and does not contain the calibration error, 
			because it will be almost canceled in the intensity ratios. 
			\label{fig:Tkin_H2CS}}
\end{figure}

The intensity ratios are compared with those calculated by using RADEX code (van der Tak et al. 2007), as shown in Figure \ref{fig:Tkin_H2CS}. 
The H$_2$ density and the column density of H$_2$CS are assumed to be from $10^7$ to $10^9$ cm$^{-3}$ and from $10^{13}$ to $10^{15}$ cm$^{-2}$, 
respectively, the latter of 
which can be compared with the previously reported value of the column density ($3.7 \times 10^{13}$ cm$^{-2}$; Blake et al. 1994) considering the beam filling factor. 
The solid and dashed lines in Figure \ref{fig:Tkin_H2CS} show the gas kinetic temperatures calculated for the various intensity ratios 
as a function of the H$_2$ densities (Figure \ref{fig:Tkin_H2CS}a) and the column densities of \hhcs\ (Figure \ref{fig:Tkin_H2CS}b). 
The gas kinetic temperatures barely depend on the H$_2$ density and the column density of \hhcs. 
Hence, they can be well evaluated from the observed intensity ratios regardless of the other parameters. 
The gas kinetic temperatures thus evaluated are listed in Table \ref{tb:temperature}. 

The gas kinetic temperatures derived above are almost consistent with the temperature at the 50 AU scale in the spherical model by Crimier et al. (2010). 
They are higher than the rotation temperatures of H$_2$CS and SO$_2$ derived from the single-dish data, 60 K and 95 K, respectively (Blake et al. 1994). 
Moreover, the gas kinetic temperatures derived from the \Kzero\ and \Ktwo\ lines of \hhcs\ seem to be higher at the centrifugal barrier than the other areas. 
As for the gas kinetic temperatures derived from the \Kzero\ and \Kfour\ lines of \hhcs, 
a similar trend of the enhanced temperature at the centrifugal barrier is seen for the red-shifted component, 
although the gas kinetic temperature derived for the Keplerian disk are higher than that derived for the centrifugal barrier for the blue-shifted component. 
These results suggest a higher gas kinetic temperature at the centrifugal barrier. 
If the dust temperature is similarly higher, 
the abundance enhancement of CH$_3$OH and HCOOCH$_3$ near the centrifugal barrier would be expected. 
If the mid-plane temperature of the Keplerian disk component is as low as 70--90 K, 
these species will be depleted onto dust grains there. 
This mechanism may be the reason why the CH$_3$OH and HCOOCH$_3$ emissions in the Keplerian disk component are not as bright as around the centrifugal barrier 
(i.e. they come from a ring-like structure around the centrifugal barrier). 
On the other hand, the deficiency of OCS and CS in the Keplerian disk component is puzzling in this context, 
because the binding energy of OCS (2888 K) and CS (1900 K) are comparable to that of \hhcs\ (2700 K) 
(UMIST database for astrochemistry; McElroy et al. 2013, http://udfa.ajmarkwick.net/index.php). 
Hence, the gas-phase destruction mechanisms of OCS and CS in the disk component have to be considered carefully. 

While the above temperature analysis supports the idea of the accretion shock for liberation of organic molecules at the centrifugal barrier, 
the temperature raise may also originate from the protostellar heating combined with the geometrical effect. 
If the infalling gas is stagnated in front of the centrifugal barrier, 
the envelope is more extended toward the direction perpendicular to the mid-plane of the envelope and becomes broader around the centrifugal barrier. 
Then, this part will be directly heated by the protostar without shielding by the mid-plane of the Keplerian disk, 
resulting in a higher gas kinetic temperature at the centrifugal barrier. 
Since the spatial resolution of the currently available data is not enough to resolve the vertical structure of the envelope, 
discrimination of the two possibilities is left for future studies. 

\subsection{Abundance of HCOOCH$_3$ Relative to CH$_3$OH} \label{sec:discuss_abundance}
The column densities of CH$_3$OH and HCOOCH$_3$ are derived 
for the envelope component, the centrifugal barrier, and the Keplerian disk component. 
Although these two molecules mainly reside in the envelope component and/or around the centrifugal barrier, 
they partly exist in the Keplerian disk (Figures \ref{fig:PV2_model}c--f). 
In order to evaluate the contribution from each component securely, 
the velocity range for integration is limited to a certain range for each component, 
as in the case of \hhcs\ (Section \ref{sec:discuss_chem}). 
Hence, the derived column densities do not mean the total ones, 
and do not have quantitative meanings by themselves. 
However, the abundance ratios HCOOCH$_3$/CH$_3$OH are meaningful for mutual comparison. 
In the evaluation of the column densities of CH$_3$OH and HCOOCH$_3$, 
we employed the LTE approximation. 
The rotation temperatures for the envelope component, the centrifugal barrier, and the Keplerian disk component 
are assumed to be 100 K, 130 K, and 100 K, respectively, 
by referring the gas kinetic temperatures derived from the \Kzero\ and \Ktwo\ lines of H$_2$CS (Table \ref{tb:temperature}). 
The derived HCOOCH$_3$/CH$_3$OH abundance ratios are shown in Table \ref{tb:columndensity}. 
 
If the \methanol\ line is assumed to be optically thin, 
the HCOOCH$_3$/CH$_3$OH abundance ratio is found to be higher than unity, 
except for the red-shifted envelope component. 
Such a high ratio is consistent with previous reports (Bottinelli et al. 2007). 
More importantly, it is lowest for the envelope component, and highest for the Keplerian disk component.  
This trend can be seen both in the red-shifted and blue-shifted components. 
It cannot simply be explained by the evaporation process, 
because the surface binding energy is almost comparable for CH$_3$OH (4930 K) and HCOOCH$_3$ (4000 K) 
(UMIST database for astrochemistry; McElroy et al. 2013, http://udfa.ajmarkwick.net/index.php). 
This may suggest that the chemical composition of grain mantles is processed to enhance HCOOCH$_3$ as grains pass across the centrifugal barrier. 
Alternatively, \MF\ may be formed in the gas phase around the centrifugal barrier and inside it (Balucani et al. 2015). 

\section{Summary} 
We analyzed the OCS, \methanol, \MF, and \hhcs\ data observed toward IRAS 16293--2422 Source A with ALMA Cycle 1 
at a sub-arcsecond resolution ($\sim 0\farcs6 \times 0\farcs5$). 
Major findings are as follows: 

\noindent
(1) The molecular distributions are different from molecule to molecule (Figure \ref{fig:PV2_obs}). 
OCS resides in the envelope, while \methanol\ and \MF\ have more compact distributions. 
On the other hand, \hhcs\ resides both in the envelope and Keplerian disk components. 

\noindent
(2) The simple infalling-rotating envelope model successfully explains the OCS distribution (Figure \ref{fig:PV18_OCS}). 
The kinematic structure of OCS in the envelope is reproduced by the model with 
the protostellar mass of 0.75 $M_\odot$ and the radius of the centrifugal barrier of 50 AU, assuming the inclination angle of 60\degr. 

\noindent
(3) The distributions of \methanol\ and \MF\ are concentrated around the centrifugal barrier (Figures \ref{fig:PV18_CH3OH} and \ref{fig:PV18_HCOOCH3}). 
They may be liberated by weak accretion shocks in front of the centrifugal barrier and/or by protostellar heating. 

\noindent
(4) \hhcs\ has high-velocity components concentrated toward the protostellar position in addition to the infalling-rotating envelope component (Figure \ref{fig:PV2_H2CS-Contour}). 
These components seem to trace the Keplerian disk component. 
Their kinematic structures can be explained by the Keplerian motion with a protostellar mass of 0.75 $M_\odot$ (Figure \ref{fig:PV2_H2CS_model}), 
which is used to explain the kinematic structure of OCS. 
Using the intensities of the \Kzero, \Ktwo, and \Kfour\ lines of \hhcs, the gas kinetic temperatures 
in the infalling-rotating envelope, the centrifugal barrier, and the Keplerian disk component 
are evaluated to be 70--120, 100--160, and 70--140 K, respectively (Table \ref{tb:temperature}). 

\noindent
(5) The \MF/\methanol\ abundance ratios are found to be 0.8--2.7, 4.2--4.9, and 8.7--8.9 
in the infalling-rotating envelope, the centrifugal barrier, and the Keplerian disk component, respectively (Table \ref{tb:columndensity}). 
The ratio tends to increase from the envelope to the Keplerian disk. 

\noindent
(6) A drastic chemical change is thus found to be occurring around the centrifugal barrier. 

The present results indicate that the centrifugal barrier plays a crucial role in hot corino chemistry.  
Recently, the existence of the centrifugal barrier has been reported for the WCCC sources (Sakai et al. 2014a, b, 2016; Oya et al. 2014, 2015), 
and therefore it seems to be a common occurrence in low-mass protostellar sources regardless of their chemical characteristics. 
Further investigations with higher angular resolution are awaited. 

\acknowledgments
The authors thank Tomoyuki Hanawa and Masahiro Machida for invaluable discussions. 
This paper makes use of the ALMA data set ADS/JAO.ALMA\#2012.1.00712.S. 
ALMA is a partnership of the ESO (representing its member states), 
the NSF (USA) and NINS (Japan), together with the NRC (Canada) and the NSC and ASIAA (Taiwan), 
in cooperation with the Republic of Chile. 
The Joint ALMA Observatory is operated by the ESO, the AUI/NRAO and the NAOJ. 
The authors are grateful to the ALMA staff for their excellent support. 
Y.O. acknowledges the JSPS fellowship. 
This study is supported by Grant-in-Aid from the Ministry of Education, Culture, Sports, Science, and Technologies of Japan (21224002, 25400223, 25108005, and 15J01610). 
N.S. and S.Y. acknowledge financial support by JSPS and MAEE under the Japan--France integrated action program (SAKURA: 25765VC).
C.C. and B.L. acknowledge the financial support by CNRS under the France--Japan action program. 

\appendix
\section{ALMA Data} \label{sec:data_ALMA}
The ALMA observations of IRAS 16293--2422 were carried out in the Cycle 1 operations 
on 22 May 2014 with the frequency setting in 230--250 GHz, 
on 14 June 2014 with the frequency setting in 220--240 GHz, 
and on 25 April 2014 with both of the frequency settings. 
Spectral lines of OCS, CH$_3$OH, HCOOCH$_3$ and H$_2$CS were observed with the Band 6 receiver at a frequency of 230, 250, 230, and 240 GHz, respectively. 
42--44 antennas were used in the observations, 
where the baseline length ranged from 17.14 to 636.53 m and from 19.58 to 628.65 m 
for the frequency settings in 230--250 GHz and 220--240 GHz, respectively. 
The field center of the observations was ($\alpha_{2000}$, $\delta_{2000}$) = ($16^{\rm h} 32^{\rm m} 22\fs72$, $-24\degr 28\arcmin 34\farcs3$). 
The primary beams (half-power beam width) are 24\farcs26 and 25\farcs06 for the observations with the frequency settings in 230--250 GHz and 220--240 GHz, respectively. 
The total on-source time were 51 minutes and 50 minutes 
for the 230--250 GHz and 220--240 GHz observations 
with typical system temperatures of 200--300 K and 50--100 K, respectively. 
The backend correlator was tuned to a resolution of 122 kHz, 
which corresponds to the velocity resolution of 0.15 km s$^{-1}$ at 240 GHz, 
and a bandwidth of 468.750 MHz. 
J1626-2951, J1700-2610 or J1625-2527 was used for the phase calibration every 8 minutes. 
The bandpass calibration was done on the following quasars: J1733-1304, J1700-2610, and J1517-2422. 
The absolute flux density scale was derived from Titan. 
The data calibration was performed in the antenna-based manner and uncertainties are less than 6\%. 
The Brigg's weighting with the robustness parameter of 0.5 was employed to obtain the images of the continuum and the spectral lines. 
In this analysis, we did not apply the self-calibration for simplicity. 
\clearpage
\begin{landscape}
\begin{table}
	\begin{center}
	\caption{Parameters of the Observed Lines 
			\label{tb:lines}}
	\vspace*{10pt} 
	\begin{tabular}{llcccc}
		\hline
		Molecule & Transition & Frequency (GHz) & $E_u$ (K) & $S\mu^2$ (Debye$^2$)\tablenotemark{a} & Synthesized Beam \\ \hline
		OCS\tablenotemark{b} & 19--18 & 231.0609934 & 111 & 9.72 & $0\farcs65 \times 0\farcs51$ (P.A. 85.\!\!\degr29) \\ 
		CH$_3$OH\tablenotemark{b} & 11$_{0, 11}$--10$_{1, 10}$; A$^{++}$ & 250.5069800 & 153 & 10.6 & $0\farcs60 \times 0\farcs47$ (P.A. 80.\!\!\degr34) \\ 
		HCOOCH$_3$\tablenotemark{c} & 19$_{9, 10}$--19$_{8, 11}$; E & 232.5972780 & 166 & 4.10 & $0\farcs64 \times 0\farcs51$ (P.A. 85.\!\!\degr78) \\ 
		H$_2$CS\tablenotemark{b} & $7_{0, 7}$--$6_{0, 6}$ & 240.2668724 & 46 & 19.0 & $0\farcs53 \times 0\farcs46$ (P.A. 73.\!\!\degr48) \\ 
		H$_2$CS\tablenotemark{b} & $7_{2, 5}$--$6_{2, 4}$ & 240.5490662 & 99 & 17.5 & $0\farcs53 \times 0\farcs46$ (P.A. 73.\!\!\degr66) \\ 
		H$_2$CS\tablenotemark{b} & $7_{4, 4}$--$6_{4, 3}$, $7_{4, 3}$--$6_{4, 2}$ & 240.3321897 & 257 &  12.8, 12.8 & $0\farcs53 \times 0\farcs46$ (P.A. 73.\!\!\degr57) \\ 
		\hline
	\end{tabular}
	\tablenotetext{a}{Nuclear spin degeneracy is not included. } 
	\tablenotetext{b}{Taken from CDMS (M\"{u}ller et al. 2005). }
	\tablenotetext{c}{Taken from JPL (Pickett et al. 1998). } 
	\end{center}
\end{table}
\end{landscape}

\clearpage
\begin{landscape}
\begin{table}
	\begin{center}
	\caption{Root Mean Squares of the Intensity Difference between the Observed and Calculated Position--Velocity Diagrams of OCS\tablenotemark{a} 
			\label{tb:goodness}}
	\begin{tabular}{cccc}
		\hline 
		& \multicolumn{3}{c}{Radius of the Centrifugal Barrier} \\ 
		Protostellar Mass \hspace*{30pt}& 30 AU\hspace*{20pt} & 50 AU & 70 AU \\ \hline
		$0.25\ M_\odot$ & 100 & 105 & 112  \\ 
		$0.50\ M_\odot$ & 88 & 91 & 98 \\ 
		$0.75\ M_\odot$ & 86 & 86 & 92  \\ 
		$1.00\ M_\odot$ & 90 & 88 & 92 \\ 
		$1.25\ M_\odot$ & 96 & 93 & 96 \\ 
		\hline
	\end{tabular}
	\tablenotetext{a}{In mJy beam$^{-1}$.} 
	\end{center}
\end{table}
\end{landscape}

\clearpage
\begin{landscape}
\begin{table}
	\begin{center}
	\caption{Gas Kinetic Temperature Derived from the H$_2$CS Lines\tablenotemark{a} 
			\label{tb:temperature}}
	\begin{tabular}{lcccccc}
		\hline 
		 & \multicolumn{3}{c}{Red-Shifted Component} & \multicolumn{3}{c}{Blue-Shifted Component} \\ 
		Transitions \hspace*{30pt} & Envelope\tablenotemark{b} & CB\tablenotemark{c} & Disk\tablenotemark{d} \hspace*{30pt} & Envelope\tablenotemark{b} & CB\tablenotemark{c} & Disk\tablenotemark{d} \\ \hline 
		\Kzero\ and \Ktwo & 70--110 & 110--140 & 70--90 & 70--110 & 100--130 & 70--120 \\ 
		\Kzero\ and \Kfour & 90--120 & 130--160 & 120--140 & 80--110 & 130--150 & $>$190 \\ 
		\hline 
	\end{tabular}
	\tablenotetext{a}{In K. 
					The gas kinetic temperatures are derived by using RADEX code (van der Tak et al. 2007). 
					The assumed ranges for the H$_2$ density and the column density of \hhcs\ are from $10^7$ to $10^9$ cm$^{-3}$ and from $10^{13}$ to $10^{15}$ cm$^{-2}$, respectively. 
					The error estimation is denoted in footnotes of Figure \ref{fig:Tkin_H2CS}. 
					} 
	\tablenotetext{b}{The infalling-rotating envelope component. The absolute value of the velocity-shift ranges from 1.0 to 4.0 km s$^{-1}$. } 
	\tablenotetext{c}{The centrifugal barrier. The absolute value of the velocity-shift ranges from 2.0 to 5.0 km s$^{-1}$. } 
	\tablenotetext{d}{The Keplerian disk component. The absolute value of the velocity-shift ranges from 5.0 to 7.3 km s$^{-1}$. } 
	\end{center}
\end{table}
\end{landscape}

\clearpage
\begin{landscape}
\begin{table}
	\begin{center}
	\caption{\MF/\methanol\ Column Density Ratios\tablenotemark{a} 
			\label{tb:columndensity}}
	\begin{tabular}{lcccccc}
		\hline 
		& \multicolumn{3}{c}{Red-Shifted Component} & \multicolumn{3}{c}{Blue-Shifted Component} \\ 
		\hspace*{50pt} & Envelope\tablenotemark{b} & CB\tablenotemark{c} & Disk\tablenotemark{d} \hspace*{50pt} & Envelope\tablenotemark{b} & CB\tablenotemark{c} & Disk\tablenotemark{d} \\ \hline
		Ratio & 0.8 $\pm$ 0.5 & 4.2 $\pm$ 0.3 & 8.7 $\pm$ 1.3 & 2.7 $\pm$ 1.2 & 4.9 $\pm$ 0.3 & 8.9 $\pm$ 1.2 \\ 
		\hline 
	\end{tabular}
	\end{center}
	\tablenotetext{a}{The quoted errors represent 3$\sigma$, where $\sigma$ is derived from the statistical error. } 
	\tablenotetext{b}{The infalling-rotating envelope. The absolute value of the velocity-shift ranges from 1.0 to 4.0 km s$^{-1}$. The rotational temperature is assumed to be 100 K. } 
	\tablenotetext{c}{The centrifugal barrier. The absolute value of the velocity-shift ranges from 2.0 to 5.0 km s$^{-1}$. The rotational temperature is assumed to be 130 K. } 
	\tablenotetext{d}{The Keplerian disk component. The absolute value of the velocity-shift ranges from 5.0 to 7.3 km s$^{-1}$. The rotational temperature is assumed to be 100 K. } 
\end{table}
\end{landscape}

\clearpage


\clearpage
\end{document}